\documentclass[aps,prb,preprint,showpacs,showkeys,groupedaddress]{revtex4}
\newcommand{\W}{\columnwidth}
\newcommand{\Wportrait}{10cm}
\newcommand{\Wtwo}{7.8cm}
\usepackage{graphicx}
\usepackage{amsmath}
\usepackage{amssymb}
\usepackage{geometry}
\newcommand{\s}{$\square$}
\newcommand{\ns}{$\not\!\square$}
\begin{document}
\title{Wetting and particle adsorption in nanoflows}
\author{German Drazer $^{a)}$}
\email{drazer@mailaps.org}
\author{Boris Khusid $^{b)}$, Joel Koplik $^{a),~c)}$, and Andreas Acrivos$^{a)}$}
\affiliation{$^{a)}$ Levich Institute, City College of the City University of New York, 
New York, NY 10031. \\
$^{b)}$Department of Mechanical Engineering, New Jersey Institute of Technology,
University Heights, Newark, NJ 07102. \\
$^{c)}$Department of Physics,
City College of the City University of New York, New York, NY 10031.}
\date{\today}
\begin{abstract}
Molecular dynamics simulations are used to study the behavior of 
closely-fitting spherical and ellipsoidal particles moving through 
a fluid-filled cylinder at nanometer scales.  The particle, the cylinder 
wall and the fluid solvent are all treated as atomic systems, and special 
attention is given to the effects of varying the wetting properties of 
the fluid. Although the modification of the solid-fluid interaction leads to 
significant changes in the microstructure of the fluid, its transport
properties are found to be the same as in bulk. Independently of the 
shape and relative size of the particle, we find two distinct regimes as a 
function of the degree of wetting, with a sharp transition between them.  
In the case of a highly-wetting suspending fluid, the particle moves through 
the cylinder with an average axial velocity in agreement with that obtained 
from the solution of the continuum Stokes equations.  
In contrast, in the case of less-wetting fluids, only the early-time motion of the particle is 
consistent with continuum dynamics. At later times, the particle is eventually 
adsorbed onto the wall and subsequently executes an intermittent stick-slip motion. 
We show that van der Walls forces are the dominant contribution to the particle
adsorption phenomenon and that depletion forces are weak enough to allow,
in the highly-wetting situation, an initially adsorbed particle to spontaneously desorb.
\end{abstract}
\pacs{47.15.Gf,47.11.+j,47.15.Rq,68.08.-p}
\keywords{nanochannel,molecular dynamics,suspension}
\maketitle

\section{Introduction}
The flow of particles of different shapes suspended in Newtonian fluids at 
low Reynolds numbers and transported through different geometries has already 
been studied for over a century, and the use of a continuum description 
through the Stokes equations to model such phenomena has proved extremely 
successful.\cite{happel} There exist, however, situations in which the flow
involves characteristic length scales far beyond the range of applicability 
of the continuum equations of fluid mechanics, and a microscopic, molecular detail 
of the fluid behavior and that of fluid-solid interfaces, may provide crucial 
understanding.\cite{KoplikB95} A paradigmatic example of this situation refers to the 
recurring question regarding the appropriate boundary conditions at the fluid-solid 
and fluid-fluid interfaces, a long standing problem that arises in the description 
of two immiscible fluids moving along a solid-surface. \cite{ThompsonR89,KoplikBW89}
In recent years, the sustained development of new microfluidic devices and their potential 
applications in a variety of fields, particularly those associated with the
so called lab-on-a-chip\cite{StoneSA04} devices or $\mu$-tas,\cite{ReyesIAM02,AurouxIRM02}
have led to a renewed interest in the problem of fluid flow under microscopic 
confinement.\cite{GravesenBJ93,HoT98,GiordanoC01,StoneK01} Moreover, novel 
fabrication techniques are reaching resolutions in the nanometer 
scale\cite{QuakeS00,LeeYMG03} and nanofluidic devices are beginning to emerge 
as powerful tools in biotechnology and related areas.\cite{HongQ03} One of 
the fundamental advantages that renders miniaturization so attractive in 
biotechnology is its potential ability to detect, transport and interrogate 
single particles at the molecular level. Of exceptional promise are those
devices that take advantage of fluid flow and suspension flow behavior only 
available at the molecular scale. It is clear then that, in order to realize 
this promise and exploit the potential benefits of such nanodevices, it is 
crucially important to understand both the transport of nanometer size particles through 
nanochannels as well as its distinctive physics under nano-confinement. In this context, 
molecular simulations provide a powerful tool that have been successfully applied to 
the study of the effects of nanoconfinement in single fluids, and have thereby illustrated 
a variety of novel and interesting dynamic behavior which drastically differs from 
the hydrodynamics of bulk fluids.\cite{HummerRN01,Grubmuller03,BecksteinS03}

In a previous letter, we investigated by means of Molecular Dynamics (MD) simulations
the low Reynolds number motion of a solid sphere in a fluid-filled capillary tube when 
the dimensions of both the particle and the channel approach the molecular scale.\cite{DrazerKAK02}
Specifically, we examined the behavior of the system as the wetting properties 
of the fluid are varied from perfect to partial wetting, and identified a novel 
adsorption phenomenon in which, for poorly wetting fluids, a solid sphere, 
initially moving along the center of the nanochannel, is adsorbed onto the tube 
wall while displacing all the fluid molecules from the particle-wall gap. 
Here we present a detailed study of such adsorption phenomena, with particular emphasis 
on:  
i)   The nature of the transition from adsorption to no-adsorption as the wetting improves, 
and its dependence on the shape of the particles; 
ii)  The role and relative importance of the depletion and dispersion forces on the 
adsorption phenomena; 
iii) The behavior of elongated particles, which exhibit orientational selectivity upon adsorption;
iv) The stochastic behavior of the adsorption phenomena, 
particularly the times at which adsorption takes place, the reversibility of the adsorption
process, and the correlation between the stick-slip motion displayed after adsorption and the 
velocity fluctuations of the particle.

\section{Molecular Dynamics Simulations}
\label{MD}

\begin{figure}
\includegraphics*[width=\Wtwo]{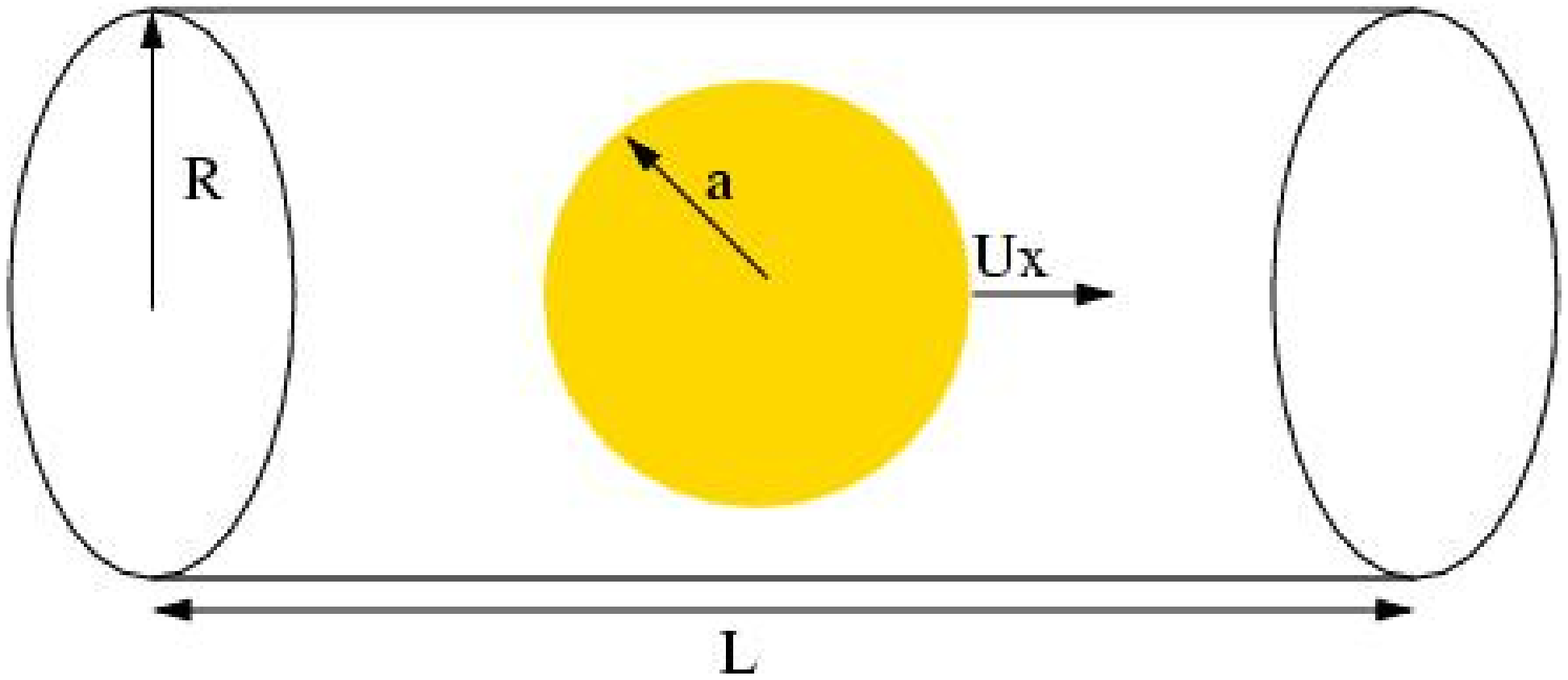}
\includegraphics*[width=\Wtwo]{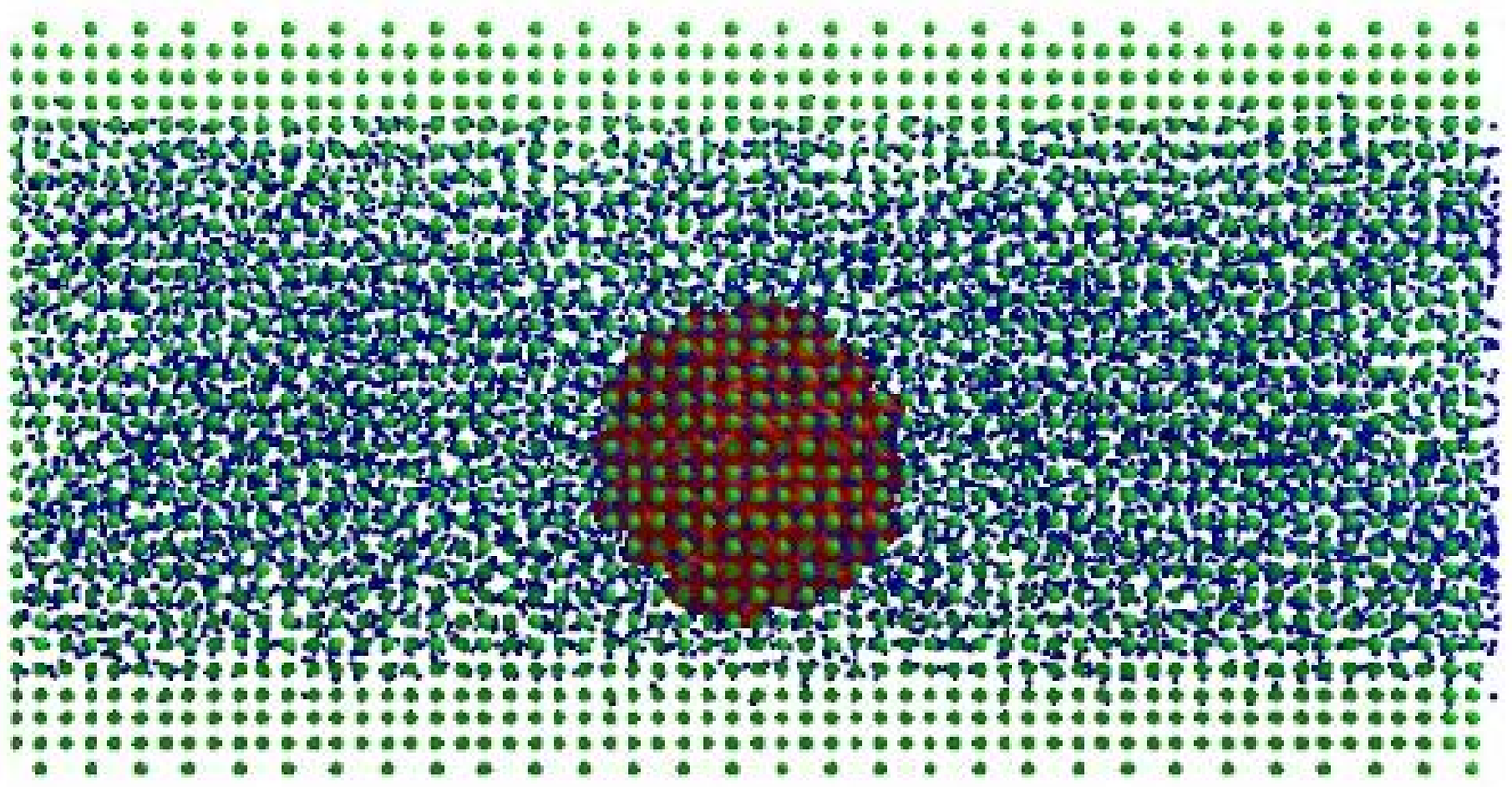}
\caption{\label{system} Schematic and explicit views of a colloidal
spherical particle moving in a nanochannel.}
\end{figure}

The numerical calculations are based on standard MD techniques applied to the 
system shown schematically in Fig.~\ref{system}, specifically, a colloidal particle 
moving under the action of an external force through a periodic cylindrical 
channel. The particle, the wall of the channel, and the fluid are all treated 
as atomic systems interacting via Lennard-Jones (LJ) potentials, 
\begin{equation}
\label{LJ}
V_{LV}=4\epsilon \left[\left(\frac{\sigma}{r}\right)^{12}-
A\left(\frac{\sigma}{r}\right)^{6}\right].
\end{equation}
Here $r$ is the interatomic separation, $\sigma$ is roughly the atomic diameter 
and will be used as a length scale, $\epsilon$ is the depth of the potential well 
and will be used as an energy scale, and $A$ is an adjustable parameter which 
controls the attraction between the various atomic species.  Although we consider 
generic non-polar materials, typically $\sigma$ is a few Angstroms, 
$\epsilon = O(10^2 k_B)$, where $k_B$ is Boltzmann's constant, and the intrinsic 
atomic time scale $\tau = \sqrt{m\sigma^2/\epsilon}$ is a few picoseconds.\cite{allen} 
In the remainder of the paper, all quantities are expressed in LJ reduced units 
using $\sigma$, $\epsilon$ and $\tau$ as characteristic length, energy and time 
scales, respectively. The potential is truncated at $r_c=2.5$ and shifted by a 
linear term so that the force vanishes smoothly at the cutoff. Newton's equations 
are integrated with a fifth-order predictor-corrector scheme, with a time step 
$\delta t = 0.005\tau$. The temperature of the system is $T=1.0$, which is maintained in 
the fluid by a Nos\'e-Hoover thermostat with relaxation time\cite{Hoover85} $Q=10$, 
and in the solid by velocity rescaling.
 
The confining wall is composed of atoms of mass $m_w=100$, tethered to fixed 
lattice sites by a stiff linear spring with constant $k_w=100$. The wall sites 
are obtained from a cylindrical section of an fcc lattice of length $L_x$ and 
inner and outer radii $R$ and $R_o=R+3\ell/2$, respectively. Here $\ell=1.71$ 
is the fcc lattice constant, corresponding to a wall number density $\rho_w=0.8$. 
We refer to this channel as an {\it ordered} cylinder, and also consider a 
{\it disordered} version, where the equilibrium  position of each atoms is 
perturbed by a random displacement in each direction, uniformly distributed in 
the range $-\ell/6<\xi<\ell/6$. 

The spherical nanoparticle is constructed in an analogous fashion, from lattice 
sites lying within a spherical volume of radius $a=3\ell=5.13$. Again we consider 
both ordered and disordered spheres, where in the latter case the atomic positions 
are perturbed randomly as is done for the wall, with $-\ell/5<\xi<\ell/5$ . 
The atoms in the particle are however fixed at the lattice sites, allowing the motion 
of the nanoparticle to be computed by rigid body dynamics.  The total force and torque 
on the particle are found by summing the individual atomic forces acting on all 
of its atoms, and Newton's and Euler's equations are integrated to give the 
particle's motion.\cite{goldstein} We assume here that the particle and channel wall 
are made of the same material and that their atoms interact via the LJ potential with $A=1$.

The fluid atoms are initialized on fcc lattice sites with the same average density 
as the wall, interacting among themselves with the $A=1$ standard LJ potential. The 
volume-average number density of the fluid is the same as that of the tube wall, 
$\rho_{av}=0.8$, which, at $T=1.0$, corresponds to the fluid phase of the Lennard-Jones 
bulk system, slightly above the critical temperature,\cite{ErringtonDT03} $T_c=0.937$.  
The fluid is allowed to ``melt'' off the original lattice sites for a few $\tau$ using 
velocity rescaling, and then is equilibrated using the Nos\'e-Hoover thermostat. A typical 
system has length $L_x=20 \ell=34.20$, inner radius $R=12 \ell=20.52$, outer radius 
$R_o=13.5 \ell = 23.09$. and contains $35860$ fluid atoms, $459$ atoms inside the 
nanoparticle, and $9920$ solid molecules in the tube wall. Figure~\ref{system} shows 
a snapshot of all atoms after equilibration.

A key feature in these simulations is the interaction between the fluid and the solid atoms, 
specified by the parameter $A$ in Eq.~(\ref{LJ}), which provides a simple means of 
controlling the wetting properties of the fluid-solid system. Recent simulations using 
similar parameters showed that varying $A$ from 1.0 to 0.5 leads to a transition from 
wetting to nonwetting behavior for a liquid drop on a solid substrate.\cite{BarratB99,BarratB99b}
Moreover, a common approximation for the contact angle in LJ systems, based on Young's 
law and a Laplace estimate for the surface energies,\cite{deGennes85,RowlinsonW} is 
$\cos(\theta)=-1+2A$. In our simulations, we vary $A$ from 0 to 1, corresponding to 
estimated contact angles ranging from $\pi$ to 0. A second interpretation of $A$ 
arises from MD simulations of dilute gas flow in the Knudsen regime, where\cite{CieplakKB00} 
$A=1$ corresponds to an ideal {\it thermal} wall in which a molecule, colliding with the 
solid, undergoes a {\it diffusive} collision and emerges from the wall with an uncorrelated  
random velocity, distributed according to a Maxwell distribution at the temperature of 
the wall. The opposite limit, $A=0$, corresponds to specular reflection where the 
normal velocity is reversed while the tangential component is preserved. 

\section{Flow of a single fluid through a cylindrical nanochannel}
\label{fluid}

\begin{figure}
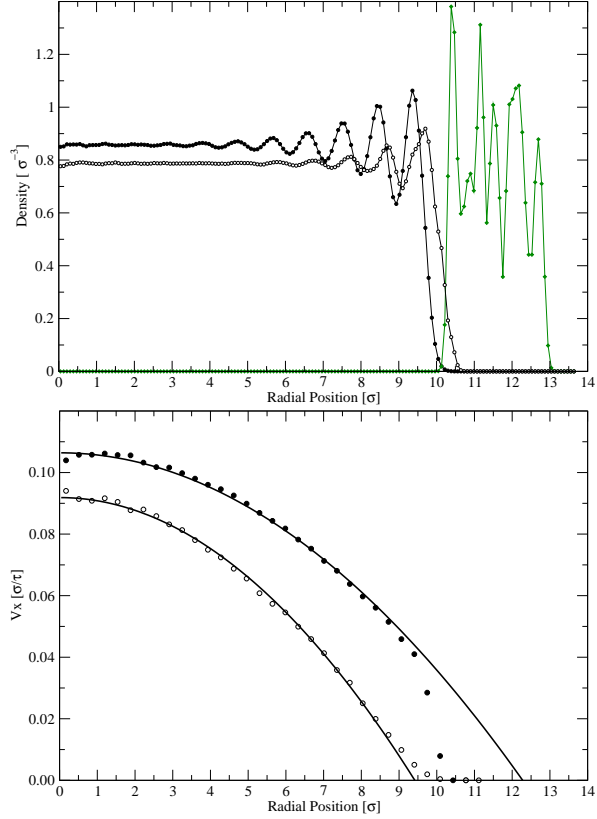

\includegraphics*[width=\Wtwo]{density_r06}
\includegraphics*[width=\Wtwo]{velocity_r06}
\caption{\label{r06} Pure fluid profiles for channel radius $R=6~\ell=
10.26$;  open and solid circles correspond to $A=1.0$ and $A=0.1$, respectively.
The data points are obtained as histograms by dividing the interior of the
tube into cylindrical shells of thickness $0.05~\ell=0.0855$.
(a) Density {\it vs}. $r$; the curves are drawn through the data points to
guide the eye.  (b) Axial velocity {\it vs}. $r$;  the curves are best fits
to a parabolic profile.}
\end{figure}

\begin{figure}
\includegraphics*[width=\W]{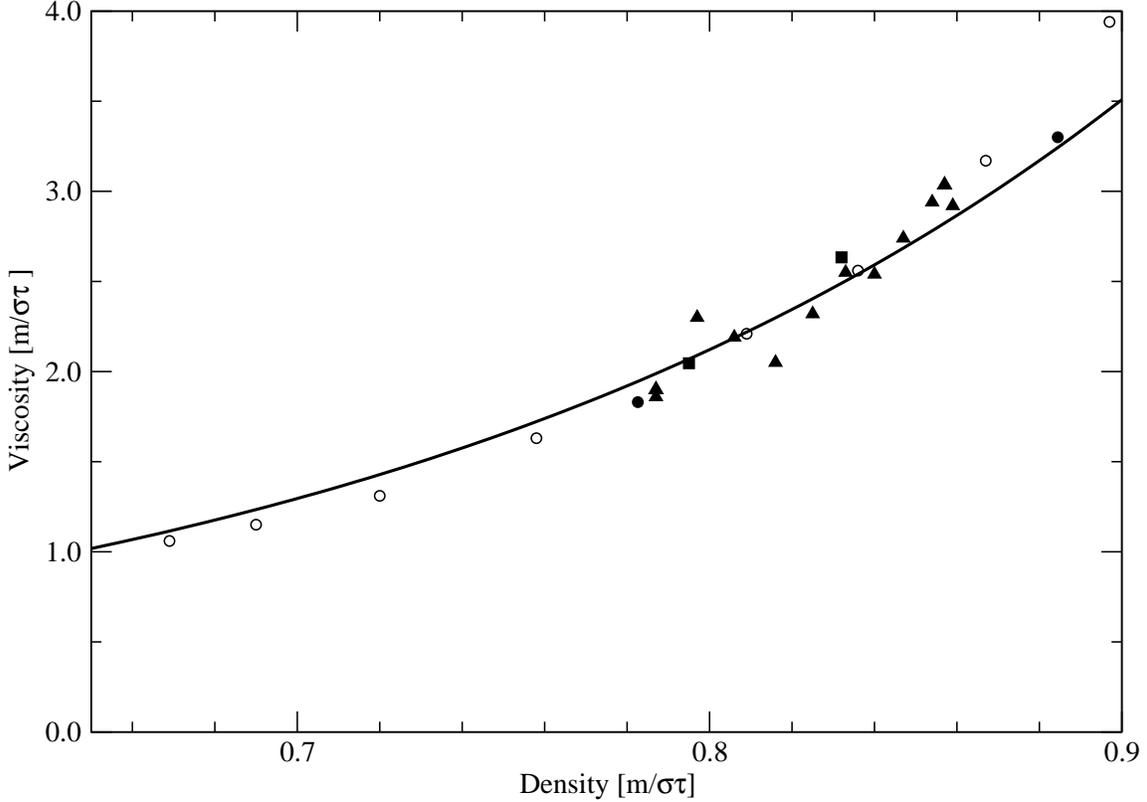}
\caption{\label{viscosity} Shear viscosity for the Lennard-Jones fluid as a function 
of the central density. The solid circles, triangles and squares are the simulation 
results listed in Table~\ref{pois} for $R=$4, 6 and 12$\ell$, respectively. The open 
circles are published numerical results for the shear viscosity of LJ bulk 
systems,\cite{Heyes88} and the solid line is an empirical fit to the experimental 
shear-viscosity data for Argon,\cite{AshurstH75} $\mu(\rho)=\mu_0 + 0.0324 (\exp(5.18 \rho)-1)$, 
where $\mu_0$ is the dilute-gas value limit of the viscosity at\cite{bird} $T=1$.}
\end{figure}

\begin{figure}
\includegraphics*[width=\W]{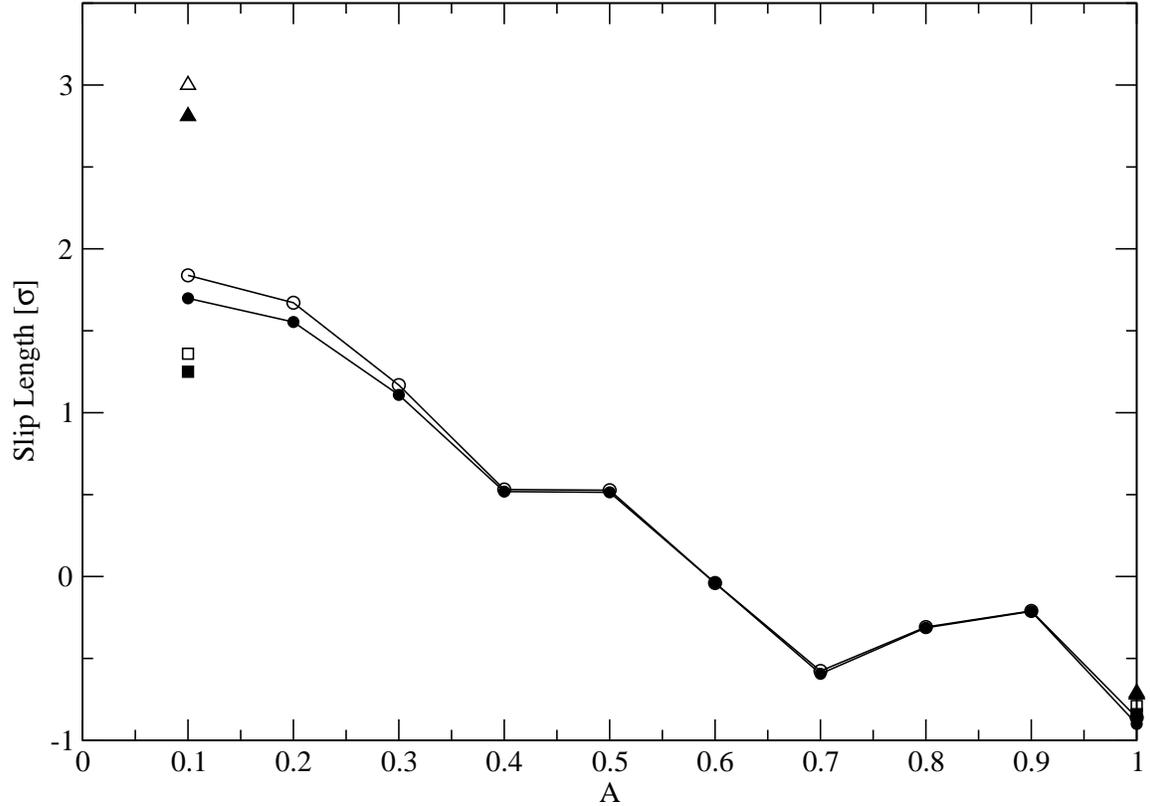}
\caption{\label{slip} Slip length for the Poiseuille flow in a narrow tube as a 
function of the wetting parameter $A$. The solid (open) squares, circles and 
triangles correspond to simulation results for the slip length $\xi_1$ ($\xi_2$) 
in tubes of radius $R=$4, 6 and 12$~\ell$, respectively.}
\end{figure}

The confinement of fluids at nanometer scales has a major impact on their 
rheological behavior, mainly due to the increasing influence of the interfaces 
as the surface-to-volume ratio is increased. We shall show, for example, that 
a variation in the solid-fluid interaction, which induces an order $\sigma$ 
displacement in the location of the first peak of the fluid density profile 
adjacent to the wall (the position of the first fluid layer that is adsorbed onto 
the wall, see Fig.~\ref{r06}) leads to noticeable changes in the average density 
at the center of the tube, in contrast to macroscopic flows. In turn, these small 
but noticeable changes in density induce substantial changes in the fluid viscosity. 
Thus, it is important to characterize first the behavior of single fluid flows in 
nanochannels under different wetting conditions, to provide some necessary 
background for the nanoparticle simulations. We begin, therefore, by discussing how 
the flow characteristics of the pure fluid vary with channel radius and wetting behavior. 
We consider Poiseuille flow, in the ordered-wall case, which is generated by applying 
to each fluid atom a body force $f$ in the axial direction. After an equilibration 
period of 5$\tau$, this force is linearly ramped up from zero over 50$\tau$, and 
then held constant at $f=0.01$ for 500$\tau$ while the time-averaged density and 
velocity profiles are measured.  We average over five realizations each for three 
values of tube (inner) radius, $R=(4,6,12)\ell$. The qualitative features of the 
results are the same for each radius, although the differences in detail merit discussion, 
so we show the profiles for the intermediate case in Fig.~\ref{r06} and some numerical 
results in Figs.~\ref{r06}-\ref{slip}, and Table~\ref{pois}. For $R=12\ell$ and $A=0.1$,
the average fluid velocity is $\bar{v}_x \sim 0.3$ and thus the Reynolds number 
is ${\rm Re}\sim 2$, which is therefore the largest Reynolds number in our simulations.

The density profiles show a wall-induced layering similar to that observed in 
liquids confined between planar surfaces, in both experiments\cite{DonnellyBAMFMMD02} 
and MD simulations,\cite{KoplikBW88,ThompsonR90} and also predicted theoretically 
for the case of hard-sphere systems.\cite{israelachvili} It is important to note 
that we have also performed equilibrium simulations in the absence of flow and 
found that the density profile agreed exactly, within the available sensitivity 
of the numerical simulations, with the non-equilibrium profiles obtained in the 
presence of flow. 
It is impressive that a system driven so far from equilibrium by extremely large 
shear rates, $\sim 10^{10}~s^{-1}$, has the same statistical structure at the 
microscopic level (singlet distribution function) as the system in thermodynamic 
equilibrium.\cite{BitsanisMTD87} In most cases the density reaches a constant 
well-defined value at the center of the tube, although its magnitude differs 
from the volume-average density $\rho_{av}=0.8$, a consequence of the attraction 
or repulsion of fluid atoms near the wall.  But, when $A=1.0$, in all cases the 
density at the center remains essentially equal to $0.8$, reflecting the fact that 
the fluid and solid have the same density and that their atoms have the same 
interactions. On the other hand, since, even for $A=1.0$, there exists still a layer 
of fluid adsorbed at the wall which slightly depletes atoms from the bulk, the 
density at the center is reduced, as seen in Table~\ref{pois}.  The number of noticeable 
layers at $A=1.0$ is 5-6.  In contrast, in simulations with poor wetting, 
$A=0.1$, the fluid avoids the wall and the first liquid layer is displaced towards 
the tube center, resulting in an obvious increase in the fluid density at the center. 
A larger number of layers is observed in this case, 6-7, probably due to the higher 
packing as a result of the same wall-repulsion effect, and, for the smallest systems, the density 
oscillations persist all the way  to the center of the channel. 

The velocity profiles in Fig.~\ref{r06} (and in the other cases not displayed) 
are also similar to those observed in LJ fluids confined between planar 
surfaces.\cite{KoplikBW88,CieplakKB01} Except for the region near the wall, 
the data are accurately fitted by a parabolic profile $v_x(r) = c_1 - c_2 r^2$, 
from which we can obtain an effective viscosity $\mu=\rho f/4c_2$ (Note that
this way of computing the effective viscosity is different from the more standard 
calculation in which the stress tensor in computed in a simple shear flow\cite{KoplikBW89}). 
In addition, a slip length can be defined either as the distance from the wall where the 
fitted velocity extrapolates to zero, $\xi_1=(c_1/c_2)^{1/2}-R$, or through 
a Navier condition relating the fluid velocity $v_s$ and the shear stress 
$\sigma_s$ at the surface, $\mu v_s = \xi_2 \sigma_s$, which gives 
$\xi_2=[v_x/(-dv_x/dr)]|_R=(c_1-c_2 R^2)/(2c_2R)$.  The two estimates of the 
slip length agree, as seen in Table~\ref{pois} and in Fig.~\ref{slip}. As was said earlier, 
in the $A=1$ wetting case, there is little variation in the fluid density at the 
center of the tube, hence only small variations in the fluid viscosity are 
observed, with mean values ranging from $\mu=1.83$ for $R=4~\ell$ to $\mu=2.05$ 
for $R=12~\ell$. On the other hand, larger variations in the central density 
for the non-wetting $A=0.1$ situation lead to even larger variations in 
the measured viscosity, which now varies from $\mu=3.3$ for $R=4~\ell$ to 
$\mu=2.63\pm0.01$ for $R=12~\ell$. However, in all cases, the viscosity variation 
correlates closely with the density variation, and agrees with previously reported 
experimental and numerical results for the shear viscosity of liquid Argon, as shown in 
Fig.~\ref{viscosity}. The slip length remains constant and negative for all tube 
sizes in the wetting simulations, $\xi\sim\sigma/2$, due to the locking of the 
first layer next to the solid surface, and again is in complete agreement with previously 
reported results in planar geometries.\cite{CieplakKB01} The non-wetting case $A=0.1$ 
shows larger slip at the fluid-solid interface, as can be seen in the velocity 
profiles, since the fluid is no longer locked to the wall,\cite{HeinbuchF89} but the 
slip length is still $O(\sigma)$, also in agreement with previous results obtained 
in planar Couette and Poiseuille flows with similar fluid densities.\cite{CieplakKB01}

\begin{table}
\caption{\label{pois}Numerical results for pure fluid flow. 
Average central density, Poiseuille fitting results, viscosity and slip lengths.}
\begin{tabular}{|l|l|l|l|l|l|} \hline
$R/\ell$& $A$ 	& $\rho(0)$ & $\mu$ &  $\xi_1$ &  $\xi_2$ \\ \hline
4	& 1.0 	&0.783 & 1.83 & -0.79&-0.84 \\ \cline{2-6}
        & 0.1 	&0.884 & 3.30 & 1.36 & 1.25 \\ \hline
6	& 1.0 	&0.787 & 1.86 & -0.86&-0.90 \\ \cline{2-6}
        & 0.9	&0.797 & 2.30 & -0.21&-0.21 \\ \cline{2-6}
        & 0.8	&0.806 & 2.19 & -0.31&-0.31 \\ \cline{2-6}
        & 0.7	&0.816 & 2.05 & -0.57&-0.59 \\ \cline{2-6}
        & 0.6	&0.825 & 2.32 & -0.04&-0.04 \\ \cline{2-6}
        & 0.5	&0.833 & 2.55 & 0.53 & 0.52 \\ \cline{2-6}
        & 0.4	&0.840 & 2.54 & 0.53 & 0.52 \\ \cline{2-6}
        & 0.3	&0.847 & 2.74 & 1.17 & 1.11 \\ \cline{2-6}
        & 0.2	&0.854 & 2.94 & 1.55 & 1.67 \\ \cline{2-6}
        & 0.1	&0.859 & 2.92 & 1.70 & 1.84 \\ \hline
12	& 1.0 	&0.794 & 2.05 &-0.71 &-0.72 \\ \cline{2-6}
        & 0.1 	&0.832 & 2.63 & 3.00 & 2.81 \\ \hline
\end{tabular}
\end{table}

\section{Motion of a sphere through a cylindrical nanochannel: Short Times}
\label{sphere}

We now consider the motion of a spherical solid particle forced
through a cylindrical tube of nanometer dimensions. As discussed
in our previous work,\cite{DrazerKAK02} the motion of the suspended
Brownian particle, initially located in the middle of the tube, can 
be divided into two different regimes. The first occurs at early 
times, for which the particle has not deviated substantially from its 
centerline motion and, thus, does not directly interact with the tube 
wall. This initial regime was shown to be amenable of a continuum description. 
The second regime correspond to the long time behavior of the particle, 
after the diffusive motion has allowed it to sample the accessible cross
section. In this section we shall examine the behavior at early times and focus on the 
particle's {\em mobility}. The ordered spherical particle of radius $a=3~\ell=5.13$ 
is initially located at the center of the tube and is allowed to equilibrate  with the 
fluid for $10\tau$, whereupon a body force is ramped up from zero to 
$f=0.1$ over a $40\tau$ interval.  We performed simulations for tube radii 
ranging from $R=4~\ell$ to $R=14~\ell$, corresponding to relative sizes 
$4/3 <R/a < 14/3$, and for a length of the (ordered) tube $L_x=30~\ell$.

\begin{figure}
\includegraphics*[width=\W]{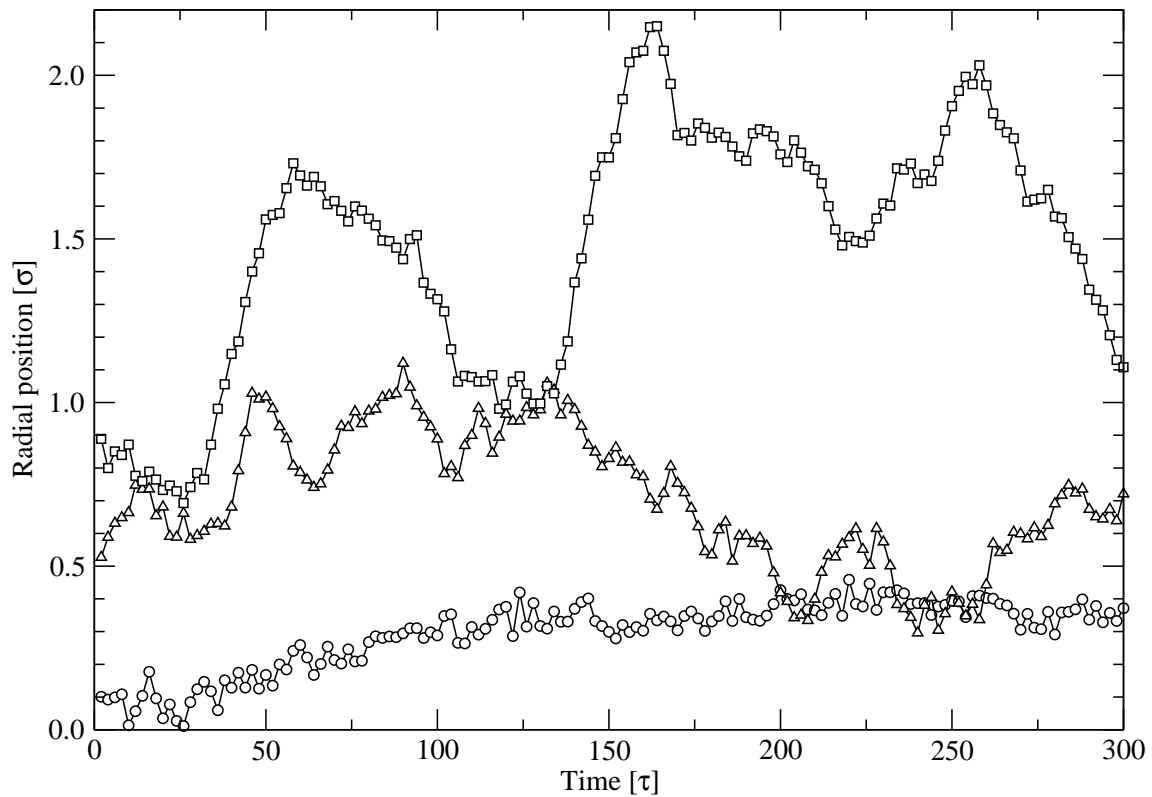}
\caption{\label{radpos} Radial position of the spherical particle, $a=3~\ell$, as 
a function of time for tube radii $R=4~\ell$ (open circles), $R=6~\ell$ 
(open triangles) and $R=14~\ell$ (open squares). Time $t=0$ corresponds to 
end of the force ramp. In all cases there is perfect wetting, $A=1.0$.}
\end{figure}

Periodic boundary conditions are used in the axial direction,
implying that the simulations in fact correspond to the motion of the
``real'' sphere and an infinite string of periodic replicas.
Since hydrodynamic interactions between suspended solid particles are usually 
long-ranged, one might be concerned about the interactions between the 
periodic images.  Fortunately, however, confining geometries effectively 
screen such hydrodynamic interactions and lead to a much faster
spatial decay of the disturbances created by a single particle, 
typically exponentially with a characteristic length fixed
by the tube radius\cite{LironS78} $R$. In particular, the interaction 
between spheres, equally spaced along the centerline of a tube, has been shown 
to be small when the separation between spheres is larger than one tube 
diameter,\cite{wangS69} and recent experiments involving colloidal particles 
in a narrow channel have demonstrated that hydrodynamic interactions are indeed 
strongly inhibited by the one-dimensional confinement.\cite{CuiDL02}  
Given that, even in the worst scenario, we have $L_x/R>2$, we can then safely 
neglect any hydrodynamic interactions with the periodic copies of the particle.

\begin{figure}
\includegraphics*[width=\W]{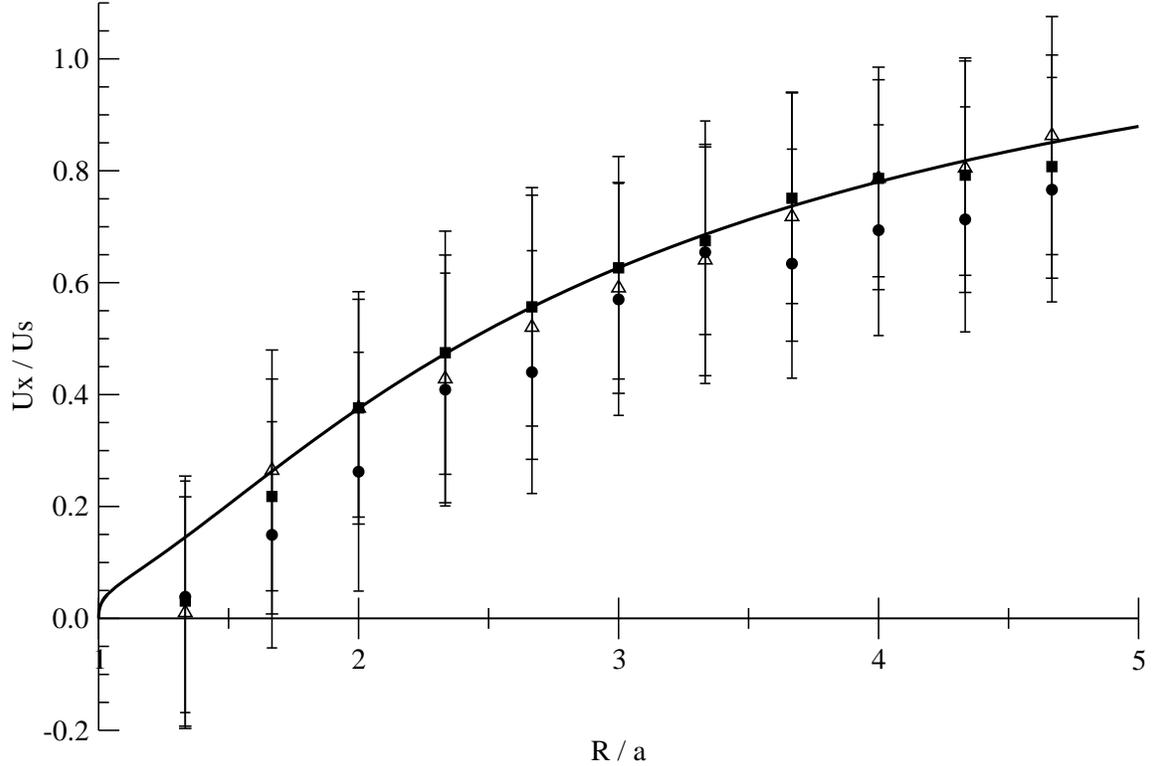}
\caption{\label{mobility} Particle average axial velocity, normalized by the Stokes velocity 
$U_s$ ($\mu_{av}=1.94$; $N_a=438$), as a function of the relative size of the tube radius: 
$A=1.0$ (closed circles), $A=0.5$ (open triangles) and $A=0.25$ (closed squares). The solid 
curve is the theoretical prediction of Ref.~\onlinecite{BungayB73}, as discussed in the text.} 
\end{figure}

\begin{figure}
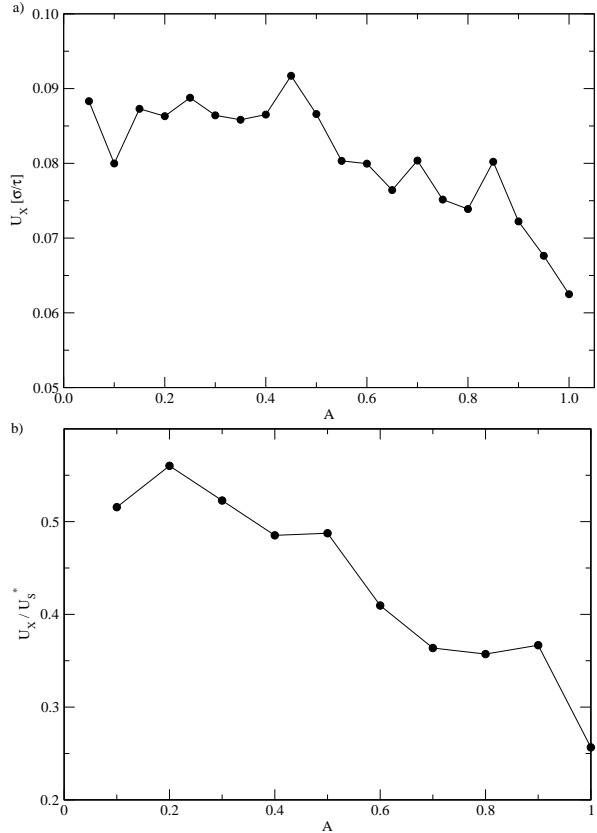

\includegraphics*[width=\Wtwo]{mobilityW}
\includegraphics*[width=\Wtwo]{mobilityWN}
\caption{\label{mobilityW} Particle average axial velocity as a function of the wetting 
properties of the system (The tube radius is $R=6~\ell$ and the particle radius is $a=3~\ell$). 
(a) The velocity is given in MD units. (b) The velocity
is normalized by a Stokes velocity $U_s^{\star}$ that accounts for 
the variations in the viscosity, $U^\star_s=F/(6 \pi \mu^{\star} a)$, 
with $\mu^{\star}$ being the viscosity values listed in Table~\ref{pois}.}
\end{figure}

The radial motion at short times is indicated in Fig.~\ref{radpos}, for
$A=1.0$ and for three different values of the tube radius.  The motion is roughly
random and diffusive, but its range in the radial direction is constrained by 
the presence of the tube, decreasing as the tube radius approaches that of the
particle.  To examine the axial velocity, we follow the evolution of the particle 
for $300\tau$, measuring the velocity every $2\tau$ and averaging the last $100$ 
points (corresponding to a time average over about half a nanosecond).
In Fig.~\ref{mobility} we present the average axial velocity of the sphere, 
$U_x$, as a function of the relative size of the tube $R/a$.  
Note the large error bars in the figure which reflect the substantial
fluctuations in the velocity of the particle due to the thermal motion.  
The axial velocity is normalized by the Stokes velocity, the velocity which the particle
would have had in an unbounded fluid under the action of the same external body force, 
$U_s=F/(6 \pi \mu a)$, where $F=N_a f$ is the total force acting
on the particle and $N_a$ the total number of atoms in it.  
Using the measured average viscosity as described in Section \ref{fluid}, 
$\mu_{av}=1.94$, and the average number of atoms in a spherical particle, 
$N_a=438$, we obtain $U_s=0.23$.  (A similar value, $U_s=0.22$, is obtained using the 
experimental value of the viscosity at the volume-average density and the 
average force on a sphere with the same density). There is some ambiguity in
the Stokes velocity arising from an ambiguity in the particle radius in that,
if instead of taking the radius of the suspended particle as $a=5.13$, we were to define 
the effective hydrodynamic radius as that of the volume inaccessible to the 
fluid molecules,\cite{VergelesKKB95} $a_{\rm eff} \approx a+1$,
the predicted Stokes velocity would decrease by about $20\%$. This uncertainty in the 
definition of the particle radius is intrinsic to its molecular nature and
the small length scales involved in the problem.
Also note that the Stokes velocity, $U_s=0.23$, was computed using the average viscosity 
obtained for $A=1.0$ and it does not account for the variations in the viscosity that occur due
to changes in the wetting properties (see Fig.~\ref{viscosity}). 
As shown in Fig.~\ref{mobilityW}(a), there is little statistically significant variation 
of the velocities with changes in the wetting parameter $A$, particularly below about 
$A=0.5$. On the other hand, when the velocity is normalized by a Stokes velocity that accounts for the 
variations in the viscosity, $U^\star_s=F/(6 \pi \mu^{\star} a)$, with $\mu^{\star}$ being the viscosity
values listed in Table~\ref{pois}, a completely different picture emerges, 
as can be observed in Fig.~\ref{mobilityW}(b), 
showing a systematic increase in the velocity with decreasing $A$.
We can then conclude that the apparent lack of dependence of the average velocity on the wetting 
properties, observed in Fig.~\ref{mobilityW}(a), is in fact due to the competition 
between an increase in the fluid viscosity and a reduction in the drag coefficient of the 
particles as $A$ is decreased.

Next, we wish to test the degree to which these results can be described by
continuum hydrodynamics.  Previous studies have shown good agreement in a
number of situations,\cite{KoplikB95} including the Stokes drag on a sphere
in an unbounded fluid,\cite{VergelesKKB96} even where the particle size is
comparable to that of the liquid molecules, and the motion of a spherical particle 
through a cylindrical pore\cite{SushkoC01} where the tube and particle 
have molecular structure and dimensions similar to those in the 
present work.  (Unfortunately, the particle's axial mobility was not considered
in Ref.\onlinecite{SushkoC01}.)  The particle's radial Brownian motion 
and molecular-scale surface roughness are difficult to address in standard 
continuum calculations, so we  compare our results with the solution of the
Stokes equations for a non-Brownian smooth sphere moving along the center 
of a smooth tube. Note that although the particle Reynolds number is $Re\approx0.5$, 
computed using the Stokes velocity of the particle in an unbounded fluid, inertia effects
will not be accounted for in the continuum calculation.
The continuum description we used also neglects density variations across the tube, 
and the associated viscosity variation discussed in Section \ref{fluid}. 
It is also assumed that the no-slip boundary condition applies at both solid-fluid 
boundaries and thus we expect this continuum description to more accurately describe 
the case $A=1.0$ for which slip effects were shown earlier to be negligible in the pure fluid 
problem in accordance with previous studies.\cite{VergelesKKB95,VergelesKKB96}

The velocity of the particle is determined only in part by the force applied 
to it, since one must also specify some additional information about the motion of
the fluid, such as  the pressure drop along the channel, $\Delta P$, or 
the mean fluid velocity along the tube, $\bar{v}_x$. Actually, it is convenient to work
with the {\it excess} pressure drop due to the presence of the sphere, 
$\Delta P_s = \Delta P - \Delta P_0$, where $\Delta P_0=8 \mu L_x \bar{v}_x / R^2$
is the Poiseuille pressure drop for the pure fluid.  The linearity
of the Stokes equations and the no-slip boundary condition imply that 
\begin{equation}
\label{resistancematrix}
\begin{bmatrix}
F \\ (\pi R^2) \Delta P_s
\end{bmatrix}
=
\mu a
\begin{bmatrix}
{\mathcal R}_{FU} & {\mathcal R}_{FV} \\ {\mathcal R}_{PU} & {\mathcal R}_{PV}
\end{bmatrix}
\begin{bmatrix}
U_x \\ \bar{v}_x
\end{bmatrix}
\end{equation}
where the resistance matrix elements ${\mathcal R}_{\mu \nu}$ are functions
of $R/a$ alone when the particle moves along the center of the tube, and were 
given explicitly by Bungay and Brenner.\cite{BungayB73,BungayB73i}

In the case at hand, we have periodic boundary conditions in the $x$ direction 
and the total pressure drop is zero. Therefore $(\pi R^2) \Delta P_s = 
- (\pi R^2) \Delta P_0 \equiv - (\mu a) {\mathcal R}_0 \bar{v}_x$, where 
${\mathcal R}_0 = 8\pi L_x/a$.
Equation~(\ref{resistancematrix}) can be solved for $F$ and $\bar{v}_x$ in terms of
$U_x$, and in particular $F=(\mu a){\mathcal R}U_x$ where
\begin{equation}
\label{bungay}
{\mathcal R} = {\mathcal R}_{FU} - 
\frac{{\mathcal R}_{FV}{\mathcal R}_{PU}}{{\mathcal R}_0+{\mathcal R}_{PV}}.
\end{equation}
In Fig.~\ref{mobility} we compare the velocity, as predicted by the above 
expression, to the numerical results, and see that the agreement is satisfactory, 
given the substantial error bars (due to thermal fluctuations).
Surprisingly, the numerical results are not sensitive to variations in
the wetting properties of the system, with a lack of apparent trend in the 
average velocity as $A$ is decreased from $A=1.0$ to $A=0.25$,
and the continuum approximations seem to describe all possible wetting conditions.
However, note that, by normalizing the particle velocity by a single Stokes 
velocity, which was computed using the average viscosity $\mu_{av}$ obtained for 
$A=1.0$, we have not eliminated the dependence on $\mu$. In particular,
its dependence on the wetting parameter $A$ is not accounted for in $U_s$. 
Thus, we can conclude again that, owing to the competition between an 
increase in the viscosity as $A$ is decreased, and the concurrent reduction 
of the drag on the particles due to a larger slip-effect, the results are 
strikingly insensitive to variations in the wetting properties of 
the fluid.

One can show, from Eq.~(\ref{bungay}) and the explicit forms of the resistance
matrix elements given in Ref.~\onlinecite{BungayB73}, that the continuum mobility 
vanishes as $(1-R/a)^{1/2}$ as $a\to R$, but, although this behavior  
is consistent with the simulations, it does not appear to capture the trend. 
Indeed, when the gap between the particle and the wall approaches a  
molecular diameter, one might expect a continuum result to fail, but we
do not have precise enough statistics to make a definitive statement. 

\section{Motion of a sphere through a cylindrical nanochannel: Long Times}
\label{secads}

The results presented thus far for the mobility of the particle correspond to 
relatively short-time motions, essentially at the center of the channel, 
where any direct interactions with the wall can be ignored. After 
a sufficiently long time, however, the thermal motion of the particle should 
cause it to sample the whole cross-section so that direct interactions with 
the tube wall are unavoidable.  An estimate of the (diffusive) time required 
to reach the wall is $\tau_D\sim (R-a)^2/2D$, where the diffusion coefficient 
in turn can be estimated from its bulk value given by the Stokes-Einstein 
formula, $D_{\rm bulk}=k_B T/6\pi\mu a$.  We find $\tau_D\sim 2500\tau$, 
roughly 10 times the elapsed time in the simulations described earlier. Therefore, 
in order to explore wall interactions, we have conducted rather longer simulations, 
$O(10^4\tau)$, in a channel of radius $R=2a$ for a range of values of $A$, 
for a {\em disordered} particle and wall.  

\begin{figure}
\includegraphics*[width=\W]{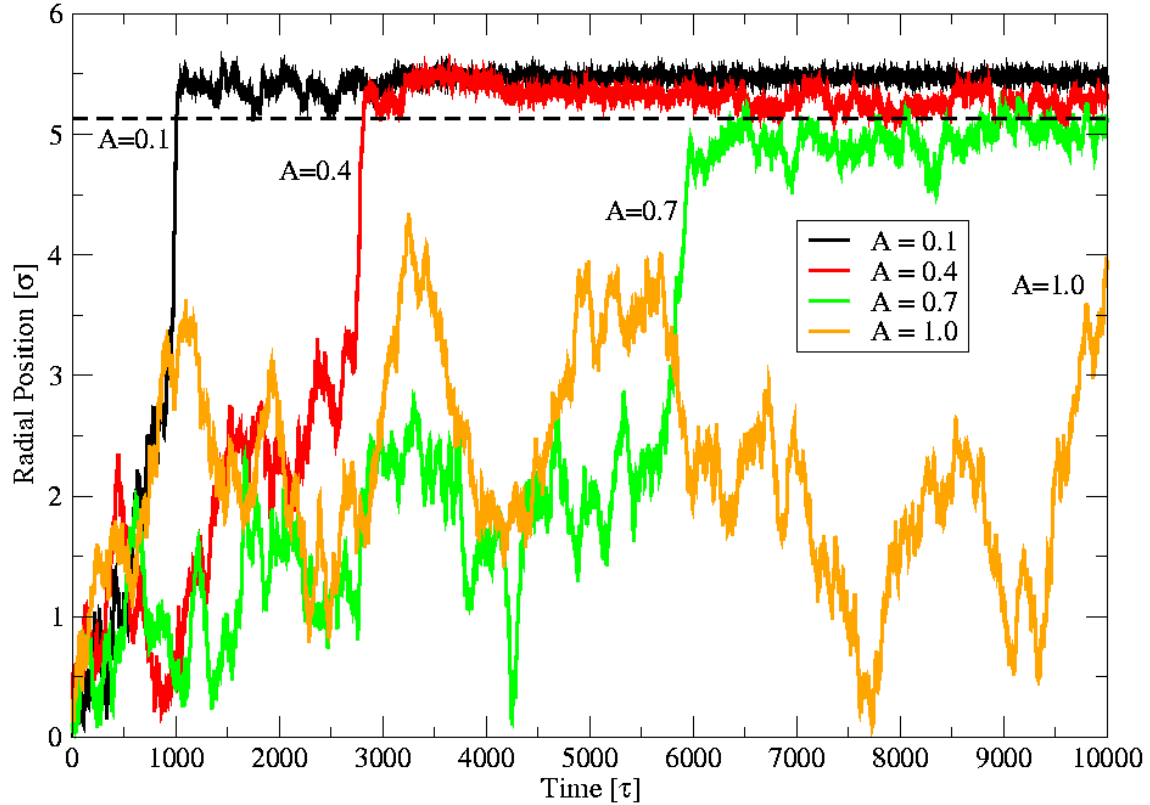}
\caption{\label{radial_pos} Radial position of the sphere {\it vs}. time 
for a single realization at four values of $A$: 0.1, 0.4, 0.7 and 1.0. 
The sudden jump to a radial postion $\sim 5$ at the three lower values of $A$ 
corresponds to adsorption at the tube wall.  
(The tube radius is $R=6~\ell$ and the particle radius is $a=3~\ell$. 
The dashed line corresponds to a radial postion equal to $R-a$, the nominal maximum radius.)}
\end{figure}

\begin{figure}
\includegraphics*[width=\W]{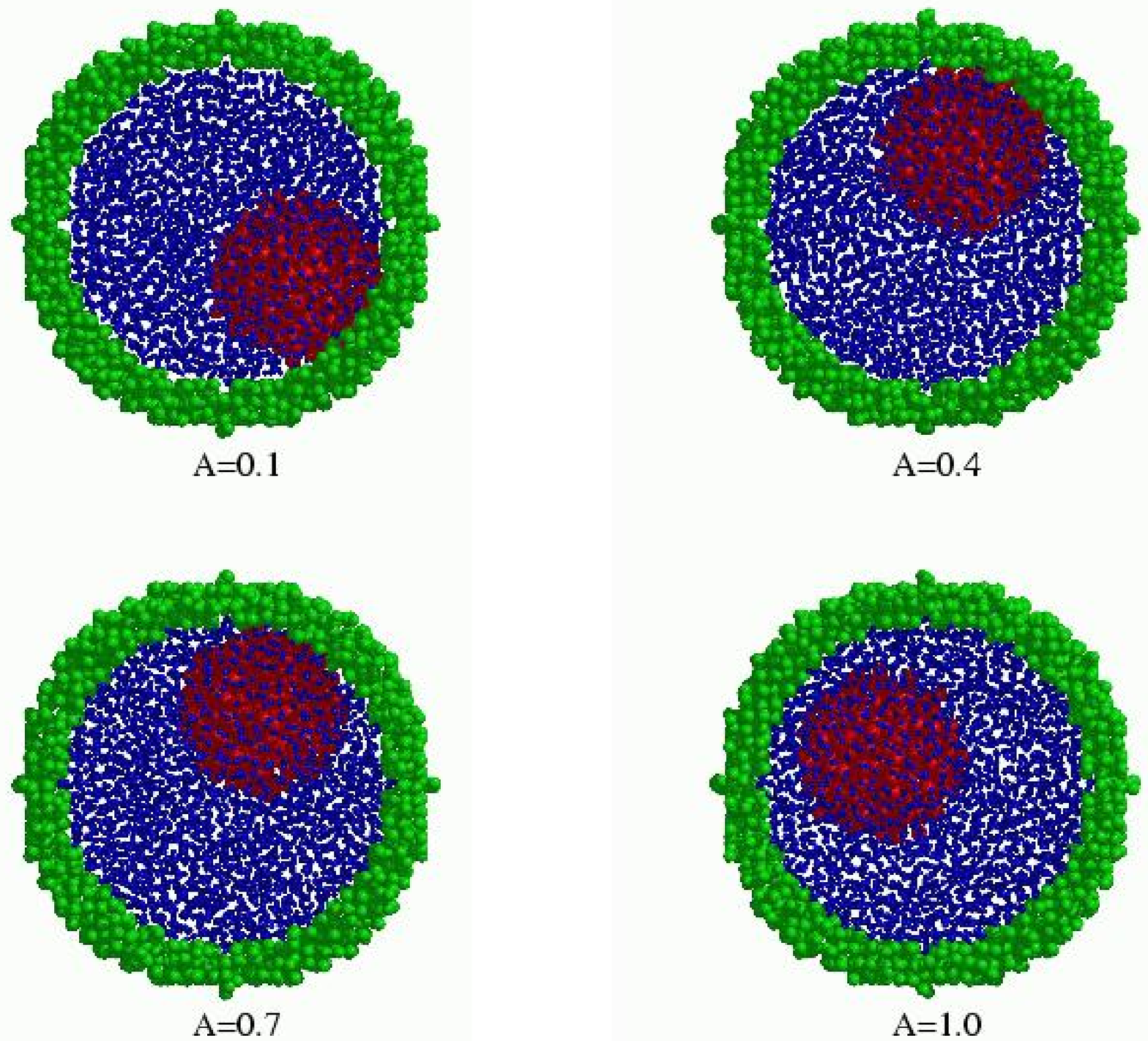}
\caption{\label{snapshots} Cross-section snapshots of the particles at 
the end of the simulations in Fig.~\ref{radial_pos}.  For visual clarity, 
we use circles of different sizes to represent the different species, with 
radii in the ratio 1.0 to 0.8 to 0.4 for the particle, wall and fluid atoms,
respectively. (The tube radius is $R=6~\ell$ and the particle radius is $a=3~\ell$.) 
} 
\end{figure}

Figure~\ref{radial_pos} illustrates the effects of the wall interaction,
showing the radial position of the particle as a function of time for four 
independent realizations in which the fluid changes from non-wetting to wetting,
corresponding to $A=0.1$, $0.4$, $0.7$ and $1.0$.  The other parameters of
the simulation are held fixed:  $R=6 \ell$; $a=3\ell$, $L_x=30\ell$ and 
$f=0.1$.  It is seen that, except for the completely wetting case, the particle eventually 
reaches a critical distance from the wall, and then rapidly moves to the
wall and becomes adsorbed, with the adsorption time decreasing with $A$. 
Note that, because the sphere and the tube are disordered and the LJ 
interaction is soft, a particle can reach a radial position with its center 
located beyond the nominal maximum value $R-a=5.13$.  The snapshots of the final 
configuration in the simulations in Fig.~\ref{snapshots} show that the adsorbed 
particles are in direct contact with the wall, having expelled the fluid from 
the gap between them, except for the completely wetting case where an intervening 
fluid layer persists. To be sure, in any continuum description to this problem, 
a lubrication force resisting the particle motion would be generated which would 
diverge as the gap between particle and wall vanishes.\cite{happel} 
As in earlier MD calculations, however, this divergence cuts off at molecular distances, 
and in fact the resistance force vanishes when all of the fluid drains out of the 
gap.\cite{VergelesKKB96}

We can obtain some insight into the particle adsorption by considering the 
problem in terms of van der Waals interactions between solids, represented by 
the attraction term of the LJ potential $-4\epsilon/r^6$. 
In a continuum description, the attraction energy between a solid wall and
a solid sphere of radius $R$, separated by an {\it empty} narrow gap of thickness 
$h\ll R$, is given by \cite{israelachvili,RusselSS,Kirsch03} 
\begin{equation}
\label{hamaker}
V_{sw}(h)=-A_H\frac{a}{6h} \left(1-\frac{a}{R} \right)^{-\frac{1}{2}},
\end{equation}
if the interaction is non-retarded and additive, where 
$A_H=4 \epsilon \pi^2 \rho_s \rho_w$ is the Hamaker constant of 
the system.\cite{israelachvili,RusselSS} (Note that in MD simulations both assumptions 
are implicit in using a constant interatomic potential.) 
The energy in this ``Derjaguin approximation'' exhibits limitations similar 
to those of the continuum description when applied at molecular 
scales in the case of the lubrication forces, in that the expression for $V_{sw}$ 
given above diverges at contact.  In fact, since the short range repulsion present in the
LJ potential would prevent molecules from approaching closer than a 
separation distance of order $\sigma$, we could use $V_{sw}(\sigma)$ as a 
reasonable estimate of the energy at contact.\cite{israelachvili}
(Note however that the snapshots in Fig.~\ref{snapshots}
show that the actual wall-sphere configuration corresponds to a more 
complex inter-meshing of the two solids, which would be rather difficult 
to capture in any continuum description.) 

\begin{figure}
\includegraphics*[width=\W]{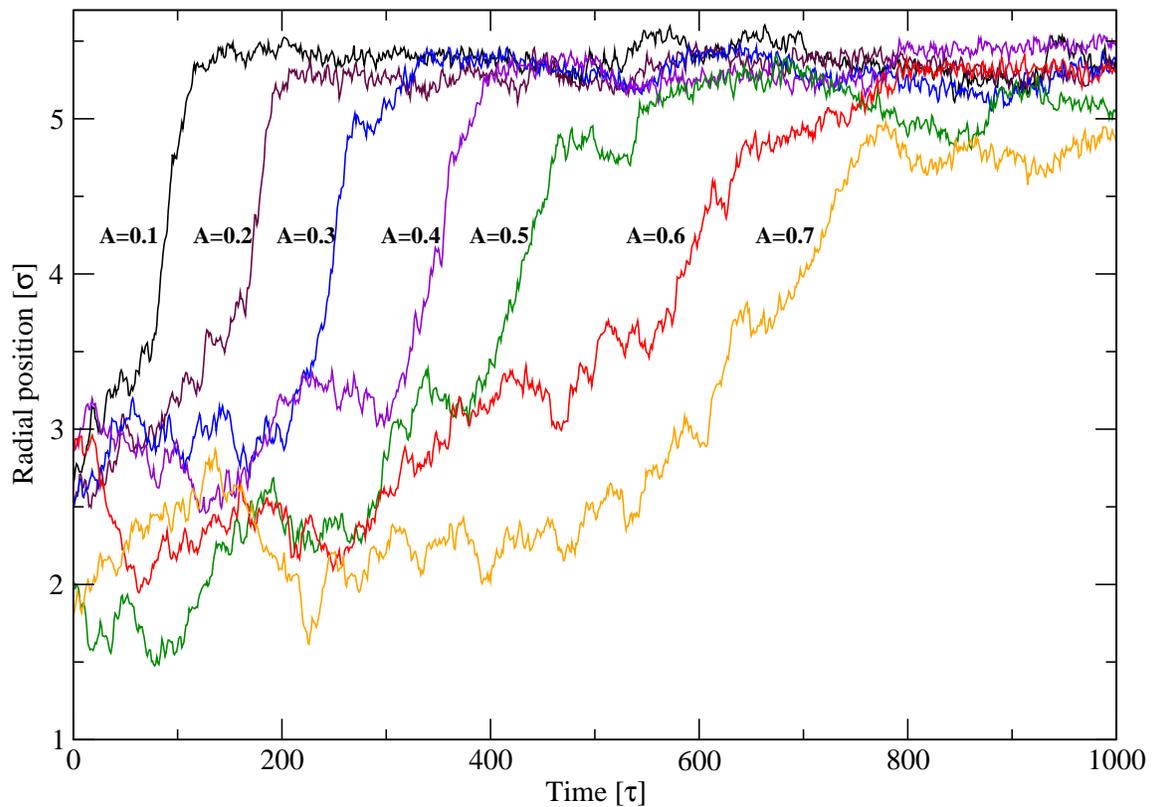}
\caption{\label{jumps} Radial position of the sphere as a function of time 
for six single realizations to $A=0.1$, $0.2$, $0.3$, $0.4$, $0.5$, $0.6$, 
and $0.7$.  The time axes are shifted so as to be able to observe
the jump into contact in all cases. 
(The tube radius is $R=6~\ell$ and the particle radius is $a=3~\ell$.)}  
\end{figure}

We can correct Eq.~(\ref{hamaker}) for the effect of the suspending fluid,
by taking advantage of the additivity of the interatomic potential 
in MD simulations. Specifically, to the expression for $V_{sw}(h)$ given above, 
we can add a term accounting for the interaction of the tube wall with the fluid 
completely filling the interior of the tube, from which we then subtract the 
interaction of the tube wall with a {\em sphere of fluid}, located at the same 
position as the solid particle. The former interaction potential is a constant, 
independent of the position of the solid sphere and can therefore be ignored, 
while the latter interaction term is of the same form as Eq.~(\ref{hamaker}) but 
multiplied by the wetting parameter $A$. Thus, the expression for the potential 
remains the same as in Eq.~(\ref{hamaker}), but with an effective Hamaker constant 
given by $A_H^{\rm eff}=4\epsilon (1-A) \pi^2 \rho_s \rho_w$.

The corresponding attractive force between the sphere and 
the tube wall is given by the derivative of Eq.~(\ref{hamaker}) with respect 
to the separation distance $h$, 
\begin{equation}
\label{force}
F_{sw}(h)=-A_H^{\rm eff} \frac{a}{6h^2} \left(1-\frac{a}{R} \right)^{-\frac{1}{2}},
\end{equation}
which is proportional to $(1-A)$.
In Fig.~\ref{jumps} we show that the adsorption transition strongly depends on $A$, 
in that the sudden jump into contact observed for $A=0.1$, becomes less abrupt
as the fluid becomes more wetting, and $1-A$ decreases, which is qualitatively consistent
with the previous result that the attractive force is proportional to $1-A$.

\begin{figure}
\includegraphics*[width=\W]{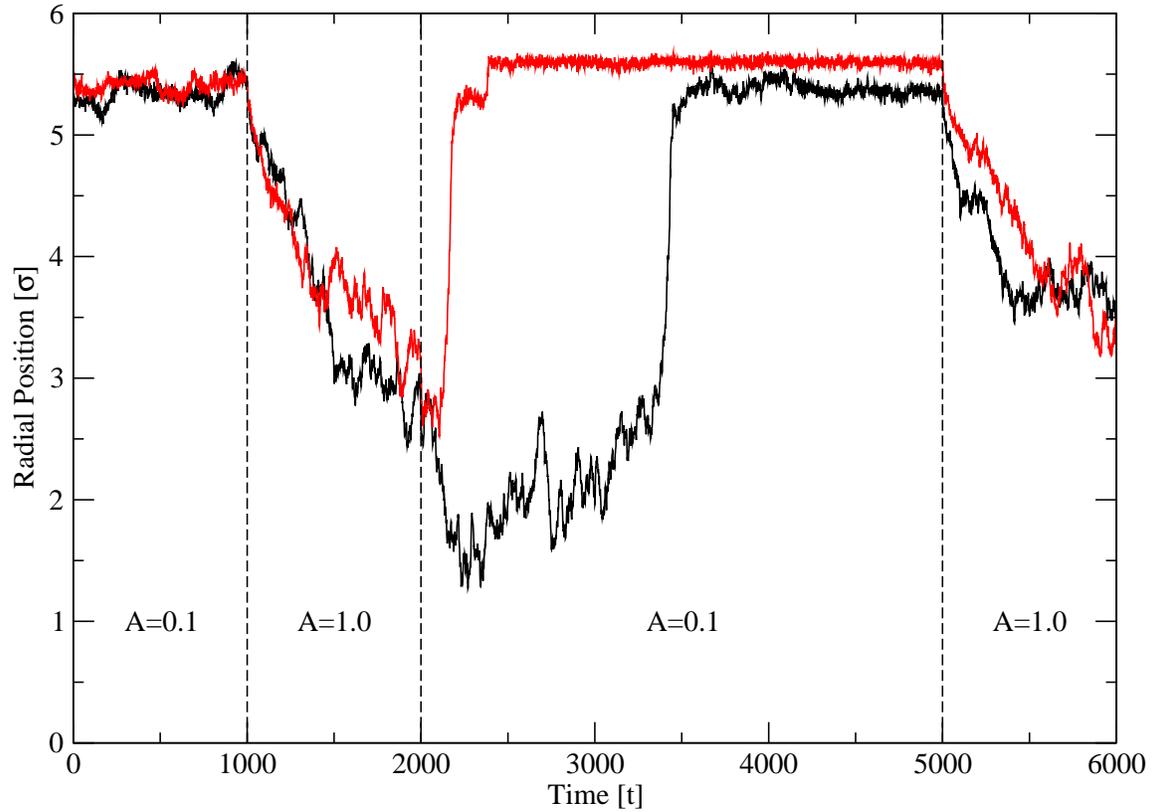}
\caption{\label{gdepletion} Radial position of the sphere as a function 
of time for two independent realizations of the MD simulations. In both 
cases the wetting parameter is set, initially, equal to $A=0.1$, leading 
to the adsorption of the particles. At $t=1000\tau$ the wetting parameter 
is changed to $A=1.0$, and a spontaneous desorption of the spheres can be 
observed. After changing, at $t=2000\tau$, the wetting parameter back to 
$A=0.1$ the spheres are re-adsorbed. (The tube radius is $R=6~\ell$ and 
the particle radius is $a=3~\ell$.)}
\end{figure}

In addition to these energetic considerations for particle adsorption, 
another possible mechanism, leading to attraction between solid bodies in a
fluid, is a depletion 
force\cite{KaplanRYP94,KaplanFL94,DinsmoreYP96,OhshimaSOTISTSKN97,DinsmoreWNY98} 
resulting from the reduction of the excluded volume and the consequent increase in
the entropy when solid particles are in close proximity.  
However, the results given previously show that changes in the interaction energy have a 
dramatic effect on adsorption, and suggest that the dominant considerations are 
energetic as further indicated from the following ``inverse'' simulations.
Specifically, we consider the evolution of an initially adsorbed colloidal particle 
suspended in a  non-wetting fluid, $A=0.1$, when the fluid is
instantaneously rendered completely wetting by changing $A$ to 1.0.
After this change, the intermolecular forces between all species are
identical, so one might expect that these forces are approximately
isotropic, and that only depletion forces would be available to preserve 
the adsorption of the particle.  
In Fig.~\ref{gdepletion} we present the evolution of the radial position 
of the sphere in two independent {\it inverse} simulations. In both cases, 
the wetting parameter is initially set at $A=0.1$ which, in time, leads 
to the sphere being adsorbed onto the tube wall; then, after allowing the sphere 
to equilibrate in an adsorbed position for $t=1000\tau$, the wetting parameter is 
switched to $A=1.0$. 
As is seen in Fig.~\ref{gdepletion}, the particles desorb spontaneously with
no measurable trapping time, thereby leading us to conclude that depletion 
forces are not responsible for the long (perhaps irreversible) adsorption times 
observed in the poor wetting cases.  This calculation is of course somewhat indirect 
in that we do not estimate directly any entropic effects, but 
in a forthcoming paper\cite{helmholtz} we calculate the Helmholtz free
energy changes associated with particle motion in nanotubes, which we hope 
will shed some light on this question.

\begin{figure}
\includegraphics*[width=\Wportrait]{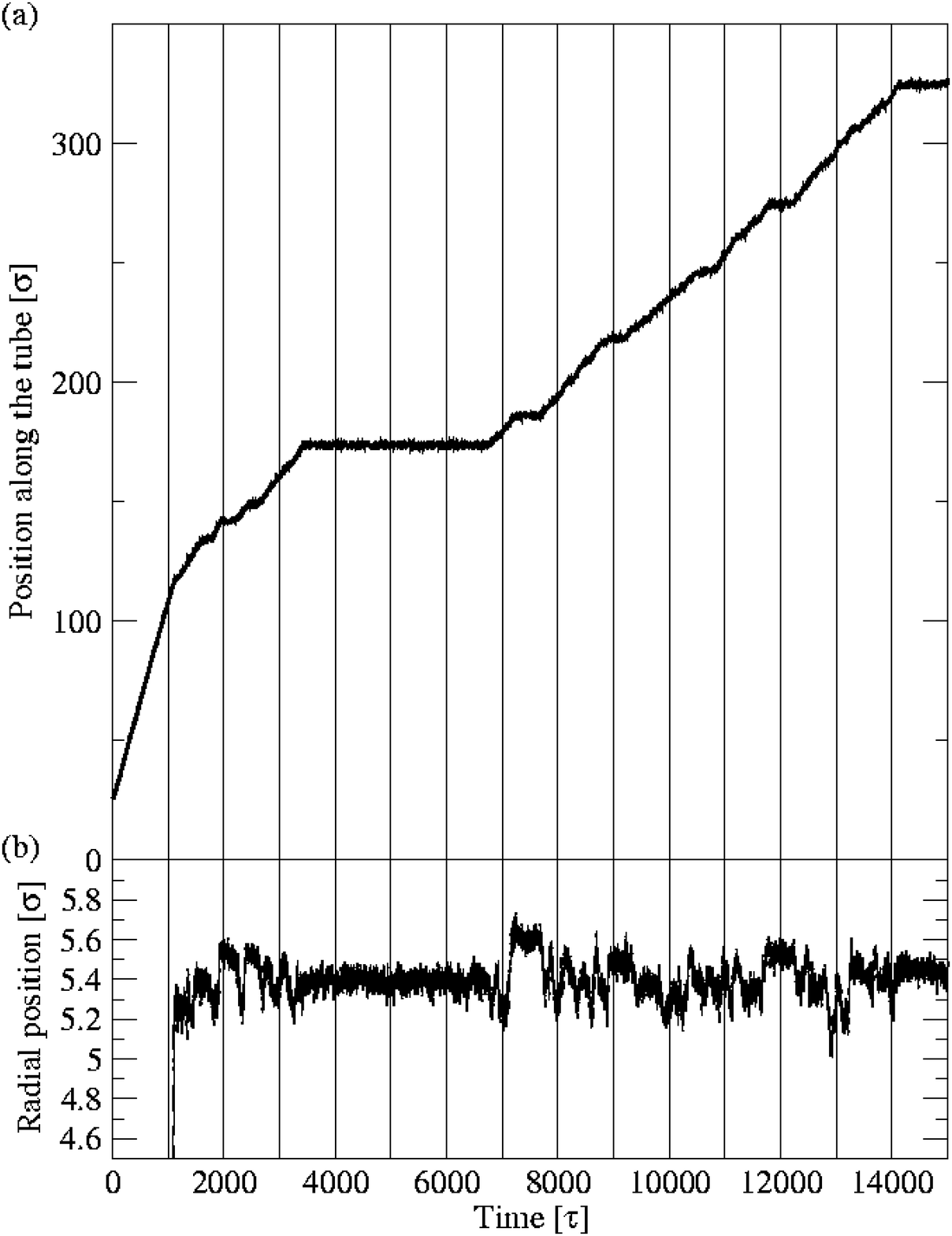}
\caption{\label{stickslip} (a) Position of the particle along the tube as a function of time
(b) Radial position of the sphere as a function of time.
The results correspond to a {\it disordered} sphere moving through a {\it disordered} tube 
for $A=0.2$. (The tube radius is $R=6~\ell$ and the particle radius is $a=3~\ell$.)}
\end{figure}

Finally, we discuss  the behavior of particles {\it after} adsorption.
In Fig.~\ref{stickslip} we present the radial and axial positions of a 
disordered sphere as it moves through a disordered tube, for 
the case $A=0.2$. It is clear that, even after adsorption has taken place, the particle
continues to move along the tube in the form of 
an intermittent {\it stick-slip} motion while remaining in near 
contact with the tube wall.  This form of a stick-slip motion has no counterpart
in any simple continuum description, because, even if it 
were possible for the sphere to overcome the resistance of the lubrication
layer of the fluid and almost touch the cylindrical wall, say at a separation $h \ll a$, the force 
required to move it along the surface would diverge as\cite{HigdonM95} $a/h$, 
which far exceeds our applied force.
Other types of interaction, such as double-layer effects, could lead to an equilibrium radial 
position of the sphere at some distance away from the wall,\cite{Adler83} 
but this also would not account for the stick-slip behavior observed here.

Stick-slip motion was found to occur for all types of ordered and disordered
spheres and walls, and thus does not depend on any precise matching of their 
underlying molecular structures.  In fact, the results
shown in Fig.~\ref{stickslip} correspond to the extreme case where
the position of both the molecules of the sphere and those of the
tube wall have been randomly modified. In our previous paper 
\cite{DrazerKAK02} we reported that, both the average radial position
of the sphere and the random fluctuations around that average, depend on 
whether the particle is sliding along the wall ({\it slip}) or whether 
it is motionless on average ({\it stick}).  For the probability density 
function of the radial position of the solid particle, we found similar
results for different types of particles and tube wall, but the distributions 
corresponding either to sliding or sticking particles have substantial 
overlap in the case of disordered solids, and the average distance to the 
wall in both cases is almost indistinguishable. On the other hand, in all
cases there exists a strong
correlation between the amplitude of the fluctuations in the radial position 
and the motion of the particle in the axial direction. 
This is illustrated in Fig.~\ref{stickslip} which shows how, during a single 
realization in which the particle displays 
intermittent stick-slip motion, the amplitude of the fluctuations 
in the radial position of the particle is clearly correlated to its axial motion.

\section{Motion of a Prolate Spheroid through a cylindrical channel}
\label{spheroids}

In practice, one may be interested in particles of more general shapes, and we 
now consider the motion of {\it prolate spheroids} through a cylindrical 
nanochannel. These spheroids were constructed in a fashion analogous to the 
spherical solid particles discussed in section \ref{MD}, with the minor axis 
equal to $a$, the radius of the spherical particles considered previously, and 
the major axis $b$ related to the ellipticity $e$ via $e=\sqrt{1-(a^2/b^2)}$. 
The simulations were performed for two different aspect ratios, $\alpha=b/a$, 
$\alpha=3$ corresponding to $e=0.89$ and $\alpha=5$ corresponding to $e=0.96$,
and both ordered and disordered spheroids were studied.
The tube radius always equaled twice the minor radius of the spheroids, $R=2a$,
and its length was changed to $L_x=50 \ell$ for the case of 
aspect ratio $\alpha=5$, in order to avoid possible hydrodynamic
interactions between periodic images.

\begin{figure}
\includegraphics*[width=\W]{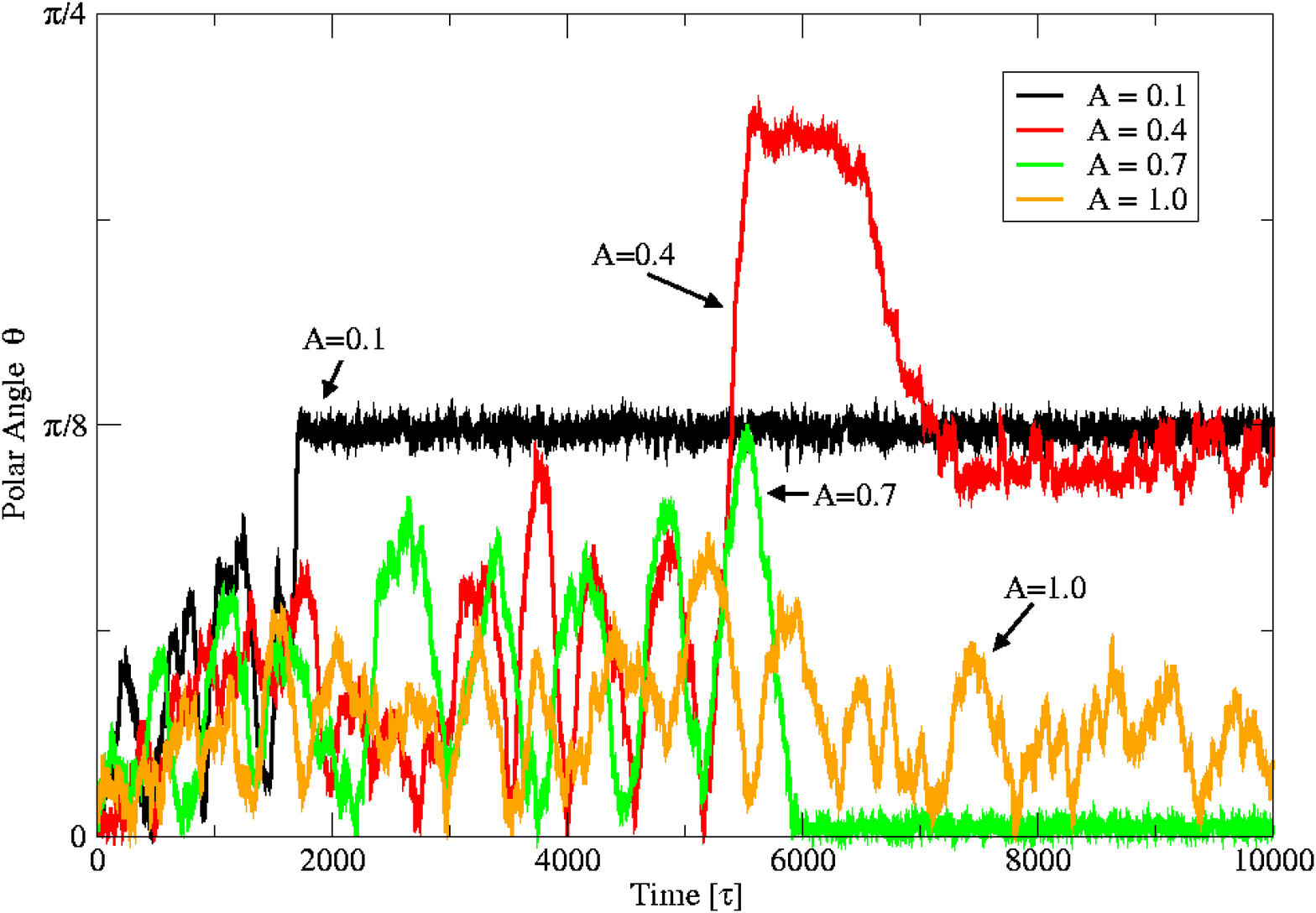}
\caption{\label{orientation} Angle between the spheroid major axis and 
the tube axis as a function of time for four realizations at
$A=0.1$, $0.4$, $0.7$ and $1.0$ (for each value of $A$ one independent MD 
simulation is presented). The abrupt locking of the polar angle 
observed at $A=0.1$, $0.4$ and $0.7$ corresponds to the adsorption of the
spheroid on the tube wall. The tube radius is $R=6\ell$. The spheroids 
have minor axis $a=3\ell$ and aspect ratio $\alpha=3$, and both the spheroids 
and the tube are ordered.}
\end{figure}

Initially the spheroids were placed at the center of the tube with their 
major axes oriented along the tube axis. Note that neither of the two 
spheroids would have fitted inside the tube if its major axis had been 
oriented perpendicular to the tube axis, with the maximum possible (polar) 
angle between the major axis of the spheroid and the tube axis being 
$\theta_m(b/a=3)\lesssim \pi/4$ and $\theta_m(b/a=5)\lesssim \pi/8$. 
As in the case of spherical particles, the surrounding fluid was first 
equilibrated and then a body force, linearly ramped up from zero to a final 
value $f=0.1$, was applied to each atom in the spheroid.  We followed the 
evolution of the spheroids as they moved through the tube for relatively 
long times, up to $10^5 \tau \sim 0.25 \mu s$ in some cases. 

\begin{figure}
\includegraphics*[width=\W]{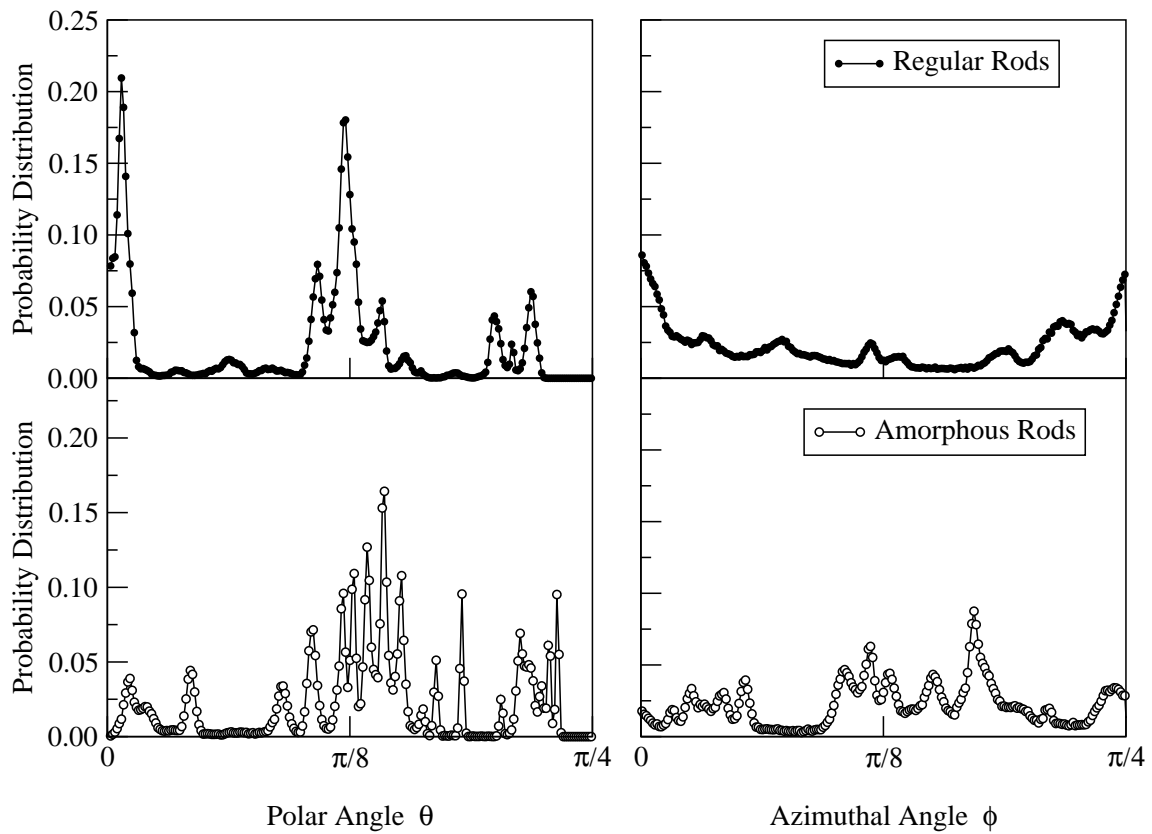}
\caption{\label{angle_pdf} Probability distribution function for the orientation 
angles of a spheroid after adsorption, averaged over 50 or more realizations with 
wetting properties ranging from $A=0.1$ to $0.7$.  The left (right) hand plots are 
the probability distribution functions for the polar angle $\theta$ (azimuthal angle $\phi$), 
and the top (bottom) plots correspond to an ordered (disordered) spheroid and wall.
The tube radius is $R=6\ell$. The spheroids have minor axis $a=3\ell$ and aspect 
ratio $\alpha=3$.}
\end{figure}
\begin{figure}
\includegraphics*[width=\Wportrait]{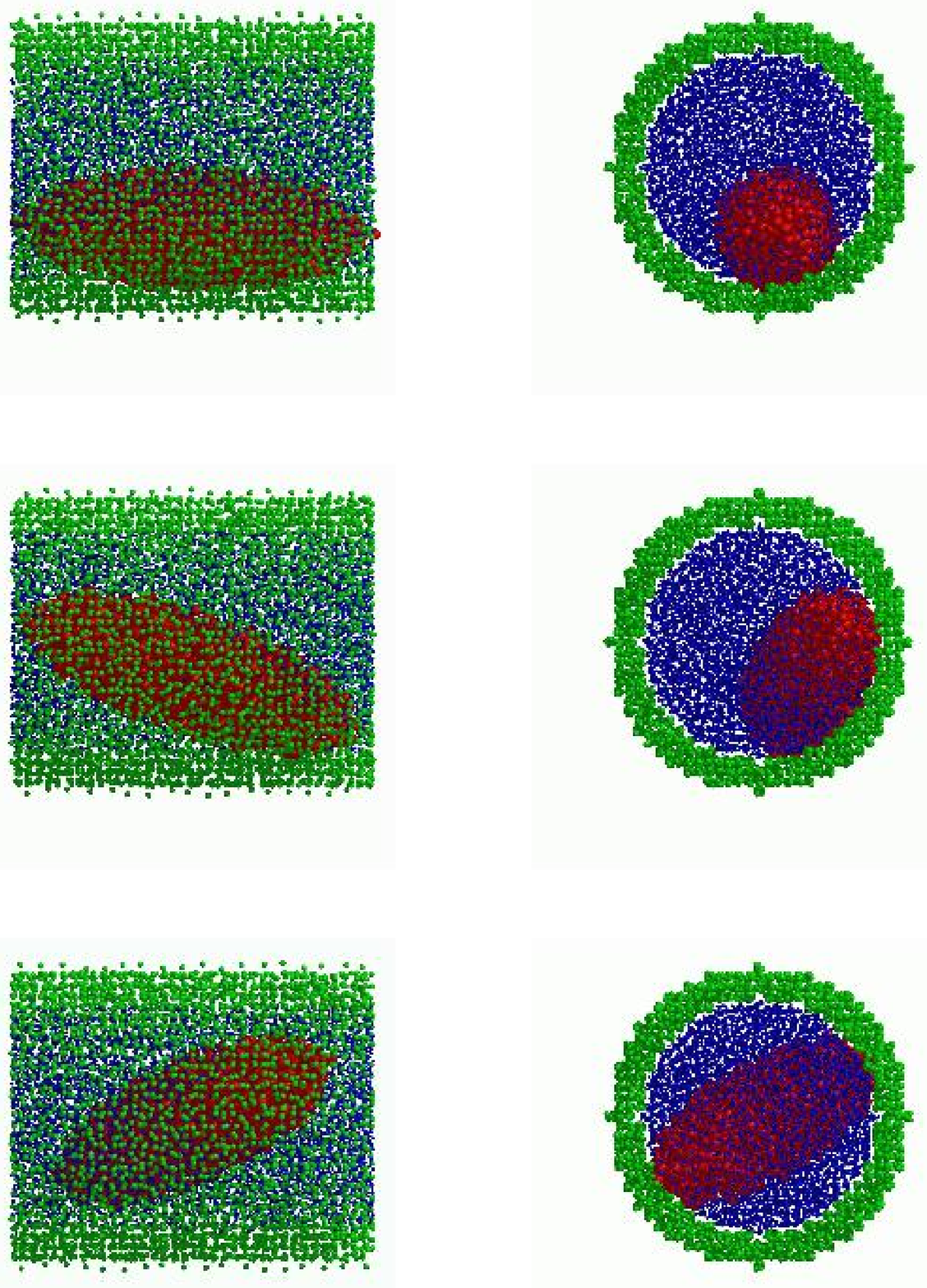}
\caption{\label{snapshots_spheroids}  Three cases of spheroid adsorption in
three distinct realizations for $A=0.1$, with side views on
the left and cross-sectional views on the right.
The tube radius is $R=6\ell$. The spheroids 
have minor axis $a=3\ell$ and aspect ratio $\alpha=3$.}
\end{figure}

The qualitative behavior of the spheroids is very similar to that of 
the spheres under similar conditions.  Specifically, in section~\ref{secads}, 
we showed that the adsorption events are readily observable in the evolution 
of the radial position, as revealed by a sudden jump-into-contact 
of the particles. In Fig.~\ref{orientation} we present the time evolution 
of the polar angle $\theta$ for independent MD simulations corresponding to different wetting 
conditions, varying from complete wetting $A=1.0$ to non-wetting $A=0.1$. 
As in the case of the sphere, for poor wetting (low $A$) the polar angle 
suddenly locks at a certain value, a transition that corresponds to the 
adsorption of the spheroid, while, in the complete wetting situation, 
the spheroid does not adsorb even after relatively lengthy simulations. 
Completely analogous results are obtained for a spheroid having a larger
aspect ratio, $\alpha=5$. 
In addition, and going beyond the observations for spherical particles, 
the spheroids display a certain selectivity of the orientation angles at 
which adsorption takes place, and this orientation remains essentially 
locked after adsorption.  In some cases, however, a transition between two
different orientations can be observed after the particle has been adsorbed
(see Fig.~\ref{orientation} for the $A=0.4$ case). 
In Fig.~\ref{angle_pdf} we present the probability distribution function 
for the orientation of the spheroids after the occurrence of adsorption, 
obtained from more than $50$ different realizations for spheroids with 
aspect ratio $\alpha=3$, and with the wetting parameter varying 
from $A=0.1$ to $A=0.7$.  We see that three different polar orientations 
predominate, corresponding to: i) spheroids with their major axis parallel 
to the tube, $b\sin\theta\approx 0$; ii) spheroids occupying one fourth of 
the tube cross section $b\sin\theta\approx R/2$; and iii) spheroids locked across the tube 
$b\sin\theta\approx R$. 
Snapshots of the three possible orientations are presented
in Fig.~\ref{snapshots_spheroids} for the case of disordered spheroids and 
a disordered tube. We can also see in this figure the complete depletion of 
fluid molecules from the narrowest gap between the tube wall and the spheroid. 

\begin{figure}
\includegraphics*[width=\Wportrait]{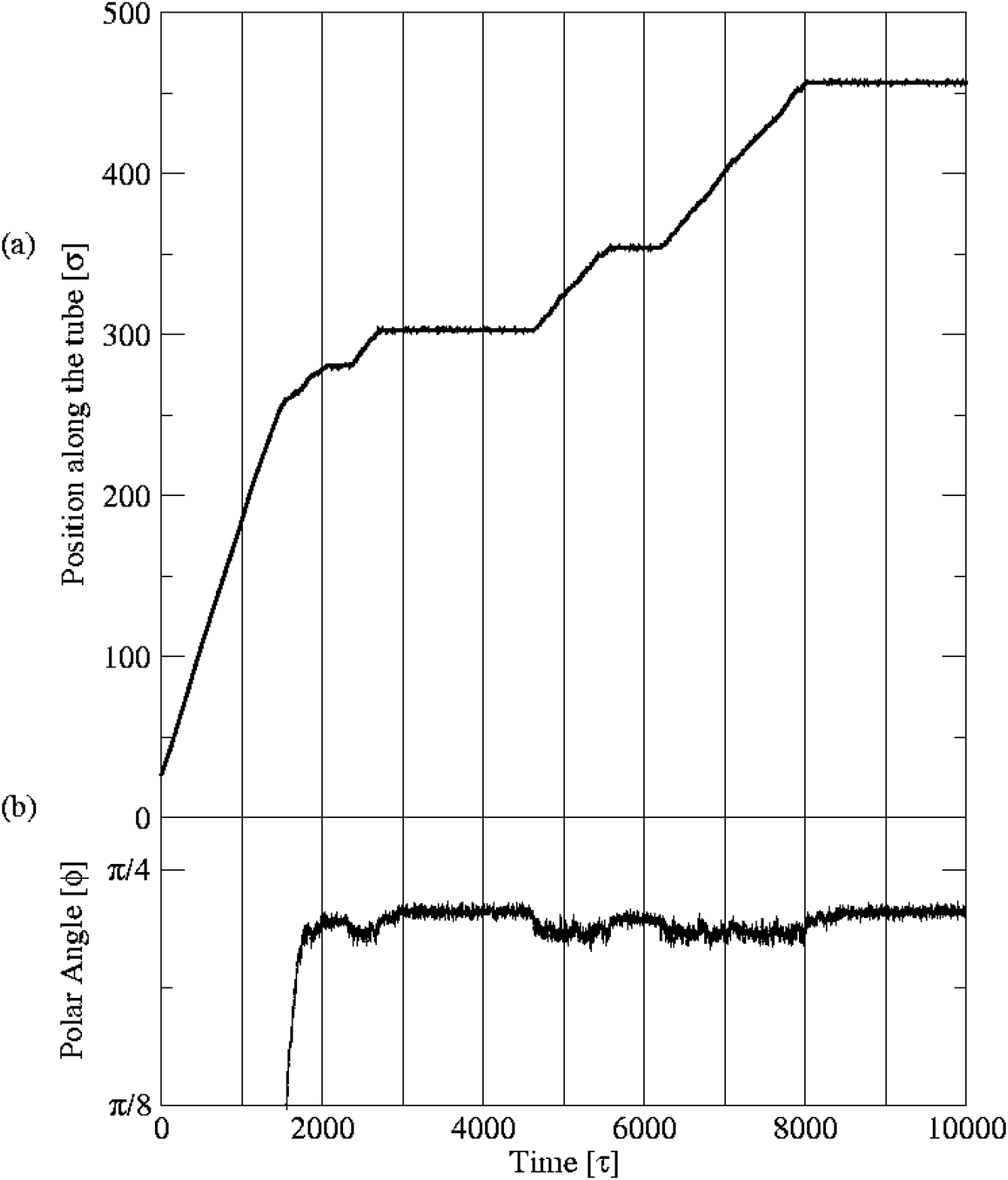}
\caption{\label{stickslipsph} (a) Position of the spheroid along the tube as a function 
of time, and (b) polar angle of the spheroid as a function of time, for a disordered 
spheroid moving through a disordered tube, for $A=0.5$. 
The tube radius is $R=6\ell$. The spheroids have minor axis $a=3\ell$ 
and aspect ratio $\alpha=3$.}
\end{figure}

Finally, we also observe that, after adsorption, the spheroids display,
in some cases, a stick-slip motion similar to that observed for adsorbed spheres. 
In Fig.~\ref{stickslipsph} we present the time evolution of the 
radial position and of the orientation for an adsorbed solid spheroid with 
$\alpha=3$ and $A=0.5$.  Clearly, there exists a strong correlation between 
the amplitude of the fluctuations in the 
polar angle and the motion of the spheroid in the axial direction, in that, 
larger angular fluctuations correspond to a sliding motion, while 
a sticking behavior corresponds to much smaller fluctuations
in the orientation.

\section{Criterion for the adsorption transition}
\label{sec.adsorption}

In the previous sections we showed that both spherical and elongated particles 
suspended in a poorly wetting fluid are eventually adsorbed onto the tube wall. 
In the simulations, the wetting properties are controlled by the parameter $A$
in Eq.~(\ref{LJ}), and a simple estimate of the attractive force between the
particle and the wall in Eq.~(\ref{force}) likewise involved $A$, so it is 
natural to ask whether there exists a critical value of $A$ which identifies an
adsorption transition.  To this end, we simulated more than 200 cases of
spheroids of aspect ratios 1, 3 or 5, in fluids with different
values of $A$, and with either ordered or disordered atomic structures for the
particle and for the tube wall, with multiple realizations in some cases.

\begin{table}\label{MDS}
\caption{Adsorption times in MD simulations. Note that an estimate of the
diffusive time required to reach the tube wall by a single sphere yields
$\tau_D\sim2500\tau$.} 
{\footnotesize
\renewcommand{\arraystretch}{0.5}
\begin{tabular}{||c|c|c|c|c|c||c|c|c|c|c|c||c|c|c|c|c|c||c|c|c|c|c|c||} \hline
A & $\alpha$ & P & C & $T_a$  &  $T_t$ &
A & $\alpha$ & P & C & $T_a$  &  $T_t$ &      
A & $\alpha$ & P & C & $T_a$  &  $T_t$ &      
A & $\alpha$ & P & C & $T_a$  &  $T_t$      \\ \hline \hline
 0.10&1&\s&\s&5	& 50& 0.66&1&\s&\s&6	& 30   &0.10&1&\ns&\s&1.5 &10   
&0.10&3&\s&\s&\{3,2,3,6,3,9,3,4,5,2\}&10 \\ \hline
 0.20&1&\s&\s&1	& 35& 0.67&1&\s&\s&2	& 30  &0.20&1&\ns&\s&2.5 &10   
&0.20&3&\s&\s&\{3,2,2,2,1,6,9,2,3,2\}&10 \\ \hline
 0.30&1&\s&\s&1	& 35& 0.68&1&\s&\s&4	& 30  &0.30&1&\ns&\s&2   &10   
&0.30&3&\s&\s&\{7,2,3,2,7,6,9,3,3\}&10   \\ \hline
 0.40&1&\s&\s&2	& 30& 0.69&1&\s&\s&12& 30  &0.40&1&\ns&\s&2.5 &10   
&0.40&3&\s&\s&\{-,-,-,3,3,6,-,7,5,4\}&10 \\ \hline
 0.50&1&\s&\s&6	& 40& 0.70&1&\s&\s&3 & 40  &0.50&1&\ns&\s&6   &10   
&0.50&3&\s&\s&\{9,5,7,5,4,3,4,-,5,7\}&10 \\ \hline
 0.55&1&\s&\s&3	& 35& 0.71&1&\s&\s&17& 30  &0.60&1&\ns&\s&-  &10   
&0.60&3&\s&\s&\{-,4,3,-,-,5,5,-,-,-\}&10 \\ \hline
 0.60&1&\s&\s&6	& 35& 0.72&1&\s&\s&6 & 30  &0.70&1&\ns&\s&4   &10   
&0.70&3&\s&\s&\{-,-,8,6,-,-,-,-,-,-\}&10 \\ \hline
 0.65&1&\s&\s&8	& 35& 0.73&1&\s&\s&11& 30  &0.80&1&\ns&\s&-   &10   
&0.80&3&\s&\s&\{-,-,-,-,-,-,-,-,-,-\}&10 \\ \hline
\multicolumn{6}{|c|}{}& 0.73&1&\s&\s&11& 30   &0.90&1&\ns&\s&-   &10   
&0.90&3&\s&\s&\{-,-,-,-,-,-,-,-,-,-\}&10 \\ \cline{7-24}
\multicolumn{6}{|c|}{}& 0.74&1&\s&\s&5	& 30   &1.00&1&\ns&\s&-   &10   
&1.00&3&\s&\s&\{-,-,-,-,-,-,-,-,-,-\}&10 \\ \cline{7-24}
\multicolumn{6}{|c|}{}& 0.75&1&\s&\s&-	& 45   &0.10&1&\ns&\ns&1  &15   
&0.10&3&\ns&\ns&\{2,3,2,1,1,4,3,3,3,8\}&10\\ \cline{7-24}
\multicolumn{6}{|c|}{}& 0.76&1&\s&\s&19& 40   &0.20&1&\ns&\ns&1  &15   
&0.30&3&\ns&\ns&\{5,1,2,2,3,2,8,3,3\}&10\\ \cline{7-24}
\multicolumn{6}{|c|}{}& 0.77&1&\s&\s&7	& 30   &0.30&1&\ns&\ns&6  &15   
&0.30&3&\ns&\ns&16&17.5\\ \cline{7-24}
\multicolumn{6}{|c|}{}& 0.78&1&\s&\s&9	& 30   &0.40&1&\ns&\ns&3  &15   
&0.50&3&\ns&\ns&\{5,2,9,8,2,5,5,8\}&10\\ \cline{7-24}
\multicolumn{6}{|c|}{}& 0.79&1&\s&\s&7	& 30   &0.50&1&\ns&\ns&1.5&15   
&0.50&3&\ns&\ns&14&15\\ \cline{7-24}
\multicolumn{6}{|c|}{}& 0.80&1&\s&\s&-	& 35  &0.60&1&\ns&\ns&4  &15   
&0.50&3&\ns&\ns&6&12.5\\ \cline{7-24}
\multicolumn{6}{|c|}{}& 0.81&1&\s&\s&-	& 15  &0.70&1&\ns&\ns&6  &15   
&0.10&5&\s&\s&\{6,3,8,6,3,7\}&10   \\ \cline{7-24}
\multicolumn{6}{|c|}{}& 0.82&1&\s&\s&-	& 15  &0.80&1&\ns&\ns&-  &15   
&0.10&5&\s&\s&11&12.5 \\ \cline{7-24}
\multicolumn{6}{|c|}{}& 0.83&1&\s&\s&-	& 15  &0.90&1&\ns&\ns&-  &15   
&0.10&5&\s&\s&\{14,14,14\}&15   \\ \cline{7-24}
\multicolumn{6}{|c|}{}& 0.84&1&\s&\s&-	& 15  &1.00&1&\ns&\ns&-  &15   
&0.30&5&\s&\s&\{12,15,3,15,5\}&17.5 \\ \cline{7-24}
\multicolumn{6}{|c|}{}& 0.85&1&\s&\s&-	& 25  &    \multicolumn{6}{c|}{}
&0.30&5&\s&\s&\{26,27\}&30 \\ \cline{7-12}\cline{19-24}
\multicolumn{6}{|c|}{}& 0.86&1&\s&\s&-	& 15  &    \multicolumn{6}{c|}{}
&0.30&5&\s&\s&21&22.5               \\ \cline{7-12}\cline{19-24}
\multicolumn{6}{|c|}{}& 0.87&1&\s&\s&-	& 15  &    \multicolumn{6}{c|}{}
&0.30&5&\s&\s&16&20\\ \cline{7-12}\cline{19-24}
\multicolumn{6}{|c|}{}& 0.88&1&\s&\s&-	& 15  &   \multicolumn{6}{c|}{} 
&0.50&5&\s&\s&\{11,29,19,19,7\}&40 \\ \cline{7-12}\cline{19-24}
\multicolumn{6}{|c|}{}& 0.89&1&\s&\s&-	& 15  &  \multicolumn{6}{c|}{}  
&0.50&5&\s&\s&80&90 \\ 		\cline{7-12}\cline{19-24}
\multicolumn{6}{|c|}{}& 0.90&1&\s&\s&-	& 25  &    \multicolumn{6}{c|}{}
&0.50&5&\s&\s&93&100 \\ \cline{7-12}\cline{19-24}
\multicolumn{6}{|c|}{}& 0.91&1&\s&\s&-	& 15  &   \multicolumn{6}{c|}{} 
&0.50&5&\s&\s&\{-,-,-\}&110 \\ \cline{7-12}\cline{19-24}
\multicolumn{6}{|c|}{}& 0.92&1&\s&\s&-	& 15  &   \multicolumn{6}{c|}{} 
&1.00&5&\s&\s&\{-,-,-,-,-,-,-,-,-,-\}&10 \\ \cline{7-12}\cline{19-24}
\multicolumn{6}{|c|}{}& 0.93&1&\s&\s&-	& 15  &   \multicolumn{6}{c|}{} 
&0.10&5&\ns&\ns&\{5,5\}&10\\\cline{7-12}\cline{19-24}
\multicolumn{6}{|c|}{}& 0.94&1&\s&\s&-	& 15  &   \multicolumn{6}{c|}{} 
&0.10&5&\ns&\ns&\{7,11,5,7\}&12.5\\\cline{7-12}\cline{19-24}
\multicolumn{6}{|c|}{}& 0.95&1&\s&\s&-	& 25  &    \multicolumn{6}{c|}{}
&0.10&5&\ns&\ns&15&17.5\\\cline{7-12}\cline{19-24}
\multicolumn{6}{|c|}{}& 0.96&1&\s&\s&-	& 15  &    \multicolumn{6}{c|}{}
&0.10&5&\ns&\ns&\{21,22,19\}&22.5\\\cline{7-12}\cline{19-24}
\multicolumn{6}{|c|}{}& 0.97&1&\s&\s&-	& 15  &    \multicolumn{6}{c|}{}
&0.30&5&\ns&\ns&\{12,12,13,8\}&15\\\cline{7-12}\cline{19-24}
\multicolumn{6}{|c|}{}& 0.98&1&\s&\s&-	& 15  &    \multicolumn{6}{c|}{}
&0.30&5&\ns&\ns&\{19,17\}&20\\\cline{7-12}\cline{19-24}
\multicolumn{6}{|c|}{}& 0.99&1&\s&\s&-	& 15  &    \multicolumn{6}{c|}{}
&0.30&5&\ns&\ns&\{17,20\}&25\\\cline{7-12}\cline{19-24}
\multicolumn{6}{|c|}{}& 1.00&1&\s&\s&-	& 25  &    \multicolumn{6}{c|}{}
&0.30&5&\ns&\ns&\{26,27\}&30\\\hline
\end{tabular}

\footnotetext{$P$: Particle type, $\square$ Ordered; $\not\!\square$ Disordered. \\
$C$: Tube wall type, $\square$ Ordered; $\not\!\square$ Disordered. \\
$T_a$: Time at which the particle is adsorbed in units of $1000\tau$;
$-$ means that there is no adsorption; $\{\}$ list of adsorption times corresponding 
to multiple realizations with identical parameters and total duration. \\
$T_t$: Duration of the simulation in units of $1000\tau$.}}
\label{MDtable}
\end{table}

\begin{figure}
\includegraphics*[width=\W]{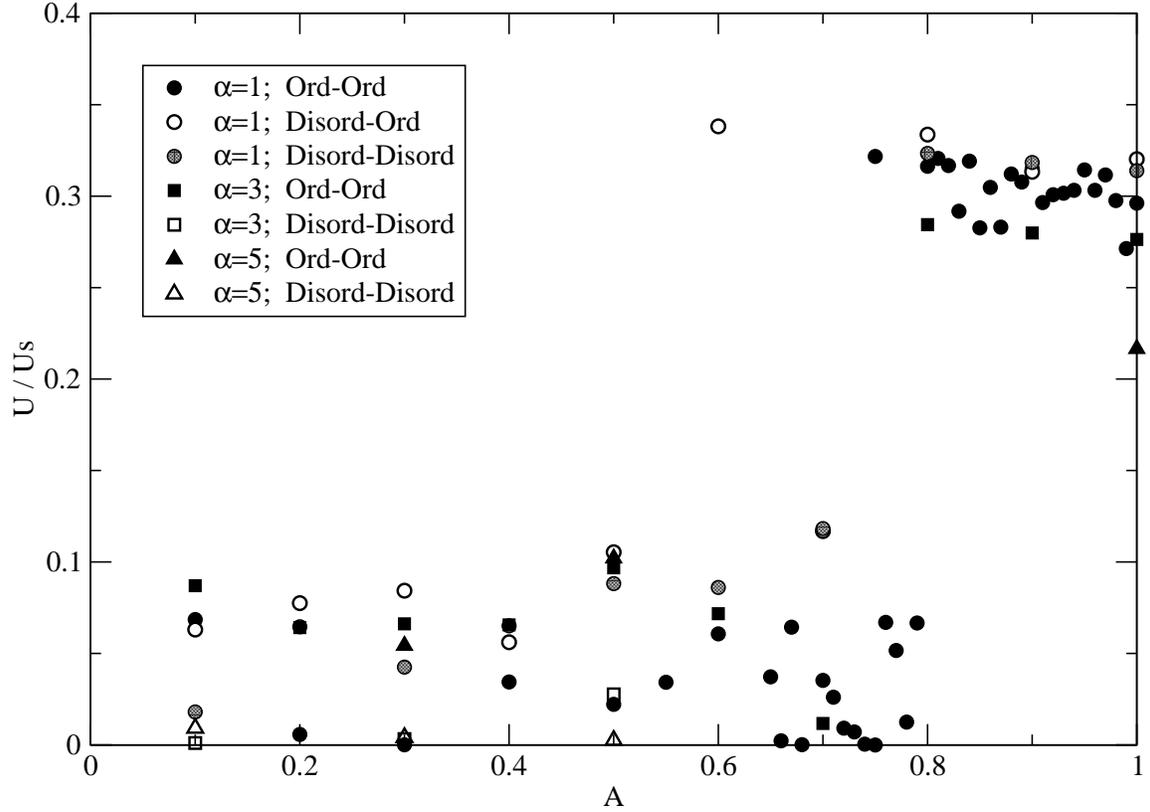}
\caption{\label{adstran} Long-time average particle velocity along the tube.
The velocity is normalized by the Stokes velocity $U_s$ that a sphere having the
same radius would have had in an unbounded fluid, for the same applied force and
no-slip boundary conditions. The time span of the simulations, $T_t \geq 10000\tau$, 
was always longer than the diffusive time required for the particle to reach the wall, 
$\tau_D\sim2500\tau$. The average velocity of particles that were adsorbed during the 
simulation is computed for times after adsorption had occurred, whereas,
for particles that are not adsorbed the velocity 
is averaged over the entire simulation. The 
dispersion in the velocities of adsorbed particles is a consequence
of the stick-slip motion after adsorption that has already been described.}
\end{figure}

Table~\ref{MDtable} summarizes the parameters of the various runs (all with
$R/a=2$).  The simulations ran for long times, $O(10^4\tau)$ or longer, 
and we can characterize the results numerically in terms of the 
average particle velocity, with the result shown in Fig.~\ref{adstran}.  
In this figure, the velocities are normalized by the 
Stokes velocity that the same particle would have had in an unbounded fluid 
with the same applied force,\cite{happel} $U_s(\alpha)=F/(6 \pi \mu a K(\alpha))$, 
where $K(1)=1$, $K(3)=1.40446$, and $K(5)=1.78481$.  Relative velocities 
of 0.1 or below correspond to the stick-slip motion of adsorbed particles already described,
while values near 0.3, which are similar to the velocities obtained at short times
({\it cf}. Fig.~\ref{mobility} at $R/a=2$),
correspond to the motion of spheres that are not adsorbed.

There exists strong evidence that the transition to adsorption occurs 
at $A\approx 0.8$ independent of particle characteristics. Moreover, we also 
performed additional ``inverse'' simulations to those discussed in Section~\ref{secads}.
We found that whereas, for $A > 0.8$, the particles are soon desorbed from the
wall, for $A \lesssim 0.8$ the particles remained adsorbed for the duration of our
simulations (e.g. $10^5\tau$ for $A=0.6$), thus providing further evidence for the existence 
of a sharp transition. Since particle adsorption is a random process which requires
a fluctuation to bring the particle near the wall, the time required can vary
over a wide range (see table~\ref{MDS}), and there is no guarantee 
that the transition to adsorption will occur during the duration of 
the simulation.  Indeed, as is seen in Fig.~\ref{adstran} there is one exception 
to the general rule, for a sphere at $A=0.6$. 

\section{Summary and Conclusions}

The complex dynamics of suspended nanometer size particles under nano-confinement 
was investigated by means of molecular dynamics simulations. We studied the low
Reynolds number transport of nanometer size spheroids with radius $a=3\ell$, and
aspect ratios ranging from $1$ to $5$, suspended in a LJ fluid of density $\rho=0.8$
and confined inside a nanochannel of radius ranging between $R=4\ell$ and $R=12\ell$.

We initially examined the behavior of single fluids flowing through a nanochannel 
for varying wetting conditions, and showed that the increase in surface-to-volume 
ratio, typical at the nanometer scales, changes the transport properties of the fluid
in a significant way. Specifically, we showed that a change in the strength of the
solid-fluid interaction leads to a small shift in the position of the first adsorbed layer of the fluid,
$\lesssim \sigma$, which in turn leads to a relatively small increase in the fluid density 
at the center the tube, $\Delta \rho/\rho \lesssim 13\%$. This small change in density, however, 
has a significant effect in the corresponding fluid viscosity at the center of the tube,
which shows an increase of up to $80\%$. However, this dependence of the viscosity on the density 
was shown to be in agreement with that observed in bulk fluids. 
The measured slip lengths are also in agreement with measurements 
obtained in bulk, or in substantially larger geometries. The structure developed by the 
fluid close to the tube wall is also consistent with previous findings as well as with equilibrium 
simulations. Finally, we showed that macroscopic continuum equations that allow for slip at 
the solid boundaries accurately describe the velocity profile of a fully developed flow in a pipe, 
even in the region where the fluid is highly structured.  

We then considered the motion of a spherical solid particle through a nanochannel, 
at both long and short times compared to the diffusive time required for the sphere 
to reach the wall. We first showed that, at short times, the average velocity of the 
particles remains essentially unaffected by changes in the wetting parameter $A$, 
particularly so for $A \leq 0.5$, and concluded that this lack of dependence on $A$ of the 
mobility of the particles is due to the competition between an increase in the 
viscosity of the fluid and the reduction of the drag force on the particles as $A$ 
decreases and the degree of wetting is reduced. Then, we compared the numerical results for the 
mobility of the spheres with the prediction based on continuum calculations for a 
non-Brownian smooth sphere moving along the centerline of a structureless tube in the 
limit of vanishingly small inertia. We found reasonable agreement, especially in view 
of the simplified continuum model that we used which does not account either for the large 
thermal fluctuations present in the system, or for the molecular structure of both 
solids. Moreover, due to the competition between drag reduction and increased viscosity 
as the degree of wetting is reduced, the continuum representation describes fairly well the 
numerical results for all wetting conditions, even though the assumed no-slip boundary 
condition becomes less accurate as $A$ decreases. At long times, however, the interactions 
of the suspended particle with the tube wall become important and we showed, that for poorly 
wetting suspending fluids, $A\lesssim 0.8$, the particles eventually are adsorbed to 
the tube wall while, surprisingly, displacing all fluid atoms from the gap, and, 
subsequently, exhibit stick-slip motion along the tube wall. 
These interesting features could never have been captured by any simple 
continuum description, where lubrication forces would prevent the particle from either 
moving towards the wall or along it for small separation distances ($h \ll a$). 
We then studied the contribution of van der Waals and depletion 
forces to the particle adsorption phenomena and concluded that the dominant contribution 
comes from the fluid-solid van der Waals interactions, as suggested by the dramatic effect 
that changes in the wetting parameter $A$ have on the adsorption process. Simulations showing 
the spontaneous desorption of particles in the complete wetting situation, $A=1.0$, 
confirmed the conclusion that depletion forces are weak enough to allow for the desorption 
of particles over extremely short times.

The behavior of prolate spheroids of two different aspect ratios, $\alpha=3$ and $\alpha=5$, 
is qualitative similar to that of the spheres, in that adsorption of the particles takes 
place for poor wetting conditions and that a similar stick-slip motion along the tube
is also subsequently observed in some cases. In addition, spheroidal particles exhibit a
certain selectivity in their final orientation upon adsorption by seemingly preferring  
three different orientations, two of which appear to locally maximize the contact area 
between the particle and the tube wall while the third appears to correspond to the 
dynamic locking of the spheroids across the tube section.

Finally, we investigated the adsorption transition as the wetting properties vary from
non-wetting to complete wetting, and found strong evidence suggesting that the transition
occurs at $A\approx0.8$ independent of particle shape. A detailed study of the Helmholtz free 
energy of the system for different wetting conditions is currently being performed and
will be the topic of a forthcoming paper.\cite{helmholtz}

\section*{Acknoledgements}

G.D. thanks J. Halverson for carefully reading the manuscript and for helpful comments.
This work was supported by the Engineering Research Program, Office of Basic Energy Sciences, 
U.S. Department of Energy under Grant DE-FG02-03ER46068. 

\bibliography{../../LaTeX/article,../../LaTeX/lanl,../../LaTeX/book,../../LaTeX/articles,../../LaTeX/mybooks}
\bibliographystyle{/usr/share/texmf/tex/latex/journals/revtex4/apsrev}

\end{document}